\documentclass[reprint,aps,prper,twocolumn]{revtex4-2} 
\usepackage{graphicx}
\usepackage{hyperref}
\usepackage{footnote}
\usepackage{amsmath,amssymb}
\usepackage{multirow}
\usepackage{tabularx}
\usepackage[table]{xcolor}
\usepackage{booktabs} 
\usepackage{longtable}
\usepackage{cleveref}
\usepackage{makecell}
\usepackage{array}
\usepackage{supertabular}
\usepackage{comment}
\usepackage{enumitem} 
\usepackage{lineno}

\usepackage{pdfpages}

\makeatletter
\AtBeginDocument{\let\LS@rot\@undefined}
\makeatother

\begin{document}


\title{Student reasoning about quantum mechanics while working with physical experiments}
\date{\today}

\author{Victoria Borish}
\email[]{victoria.borish@colorado.edu}
\author{H. J. Lewandowski}

\affiliation{Department of Physics, University of Colorado, Boulder, Colorado 80309, USA}
\affiliation{JILA, National Institute of Standards and Technology and University of Colorado, Boulder, Colorado 80309, USA}

\begin{abstract}
Instruction in quantum mechanics is becoming increasingly important as the field is not only a key part of modern physics research, but is also important for emerging technologies. However, many students regard quantum mechanics as a particularly challenging subject, in part because it is considered very mathematical and abstract. One potential way to help students understand and contextualize unintuitive quantum ideas is to provide them opportunities to work with physical apparatus demonstrating these phenomena. In order to understand how working with quantum experiments affects students' reasoning, we performed think-aloud lab sessions of two pairs of students as they worked through a sequence of quantum optics experiments that demonstrated particle-wave duality of photons. Analyzing the in-the-moment student thinking allowed us to identify the resources students activated while reasoning through the experimental evidence of single-photon interference, as well as student ideas about what parts of the experiments were quantum versus classical. This work will aid instructors in helping their students construct an understanding of these topics from their own ideas and motivate future investigations into the use of hands-on opportunities to facilitate student learning about quantum mechanics.
\end{abstract}

\maketitle

\section{Introduction}\label{sec:intro}

Quantum mechanics, as one of the foundations of modern physics with many technological applications, is an important part of physics students' education. However, its abstractness and lack of relevance to students' everyday experiences can make it particularly challenging for students to learn \cite{johnston1998student, singh2009cognitive, bouchee2022towards}. One way to provide students a  concrete context to consider as they learn about unintuitive quantum phenomena is the use of quantum optics experiments, such as ones that demonstrate particle-wave duality of photons. These experiments have been incorporated into courses as thought experiments \cite{baily2012interpretive, kohnle2014investigating, malgieri2014teaching}, simulations \cite{kohnle2014investigating, malgieri2014teaching, mckagan2008developing, kohnle2015enhancing, marshman2022quilts}, interactive screen experiments \cite{bronner2009interactive, bitzenbauer2021effect}, and physical experiments \cite{galvez2005interference,beck2012quantum,lukishova2022fifteen,borish2023implementation}. 

The use of physical quantum optics experiments in undergraduate courses has become increasingly popular over the past two decades \cite{galvez2005interference,beck2012quantum,lukishova2022fifteen,borish2023implementation}. Many of these experiments, which are commonly called the ``single-photon experiments,'' utilize heralded and entangled photons to demonstrate foundational topics in quantum mechanics, such as single-photon interference \cite{galvez2005interference, beck2012quantum} or a violation of local realism \cite{dehlinger2002entangled,carlson2006quantum}. These experiments have been disseminated throughout the advanced labs community in the United States, in part through yearly workshops aimed at teaching instructors how to set up and incorporate the experiments in their own courses \cite{alphaWebsite}. Instructors choose to use the single-photon experiments for various reasons, including helping students learn concepts about the wave-like nature of photons and understand the differences between quantum and classical models of light \cite{borish2023implementation}.

Although working with physical experiments may provide students unique opportunities to make sense of quantum phenomena differently than they could without access to the experimental apparatus, there have been relatively few studies investigating the effectiveness of this approach. Students have been shown to report learning concepts \cite{galvez2010qubit, borish2023seeing} and score better on assessments \cite{lukishova2022fifteen} after working with the single-photon experiments, yet we are not aware of any prior research investigating how productive student reasoning develops as students interact with the experiments.

In the work presented here, we investigated how physical experiments may help students improve their conceptual understanding of quantum mechanics by analyzing the in-the-moment student thinking as students worked with the single-photon experiments. We recorded lab sessions of two pairs of students outside of a course context as they worked through a sequence of three experiments that demonstrated both particle-like and wave-like behavior of light. By analyzing the students' conversations while they worked with the experiments, as well as their reflections on the process afterward, we were able to answer the following research questions:
\begin{enumerate} [label=RQ\arabic*.]
\itemsep0em
    \item What resources do students activate when confronting experimental evidence of single-photon interference?
    \item What parts of the single-photon experiments do students identify as quantum or classical, and how does this change as they work with the experiments?
\end{enumerate}
This is one of the first studies investigating student reasoning throughout such a complex experiment, and it provides insight into how students understand not only their experimental results but also what it means for something to be quantum. 

We present our results to these two research questions separately, beginning with some relevant background information. In Sec.~\ref{sec:background}, we provide more details about the single-photon experiments, the resources framework underlying our first research question, and prior research on student conceptual learning of quantum mechanics through an experimental context with or without a physical apparatus. Then, in Sec.~\ref{sec:methods}, we expand on our methods including details of the three experiments the students in our study performed, the data we obtained, our analysis methods, and the limitations of our study. Next, we present and discuss our results to RQ1 in Sec.~\ref{sec:resources} and RQ2 in Sec.~\ref{sec:quantumVsClassical}. In both of these sections, we first present conversations and quotes from the students and follow that with a discussion that synthesizes the student ideas into takeaways for instructors. We conclude in Sec.~\ref{sec:conclusions} by discussing some connections between the two research questions and future research directions suggested by this work.

\section{Background} \label{sec:background}

In this section, we provide additional context for our study, beginning with a description of the single-photon experiments. Since we are investigating student conceptual learning while working with these experiments, we then discuss the resources framework that we used to define and answer RQ1. This is followed by a discussion of some prior research on students' conceptual understanding of particle-wave duality of photons to allow for connections between our findings and already-known difficulties students have when learning about these topics without the experimental apparatus. 

\subsection{Single-photon experiments}

The single-photon experiments are a set of quantum optics experiments that have become popular within undergraduate courses \cite{dehlinger2002entangled,thorn2004observing,galvez2005interference,pearson2010hands,beck2012quantum,lukishova2022fifteen}. Some of these experiments were first implemented in undergraduate courses in the early 2000's \cite{dehlinger2002entangled,holbrow2002photon,thorn2004observing,galvez2005interference,gogo2005comparing,pysher2005nonlocal} and since then they have spread to many other institutions with new variations continuing to be developed for educational purposes \cite{bronner2009demonstrating,ashby2016delayed,galvez2023curriculum}. In the United States, the popularity of these experiments has been at least in part facilitated by the Advanced Laboratory Physics Association, which sponsors yearly workshops teaching instructors how to implement these experiments and helps instructors buy the necessary equipment at a discounted price \cite{alphaWebsite}.

The single-photon experiments are currently used in undergraduate courses in a variety of ways to introduce students experimentally to topics such as particle-wave duality of photons and entanglement. They are most often incorporated into either upper-division quantum mechanics or beyond-first-year (BFY) lab courses. When used in BFY lab courses, the students may or may not have already taken a quantum mechanics course. Instructors hope these experiments will help their students accomplish many different learning goals ranging from learning quantum concepts to improving lab skills to increasing motivation for coursework and future research. Based on their goals, instructors choose to incorporate one or several of these experiments, which all use a similar apparatus \cite{borish2023implementation}. 

In this work, we focus on three of the individual experiments that are commonly used together to demonstrate particle-wave duality of photons. In the first experiment, students set up  detectors that measure pairs of entangled photons created by the process of spontaneous parametric down-conversion (SPDC) \cite{beck2012quantum}. In the other two experiments, students use these pairs of entangled photons as a heralded single-photon source, so a photon hitting one detector indicates the existence of its partner photon in a single-photon state. This heralded single photon can be sent through a beam splitter to show that the single photon can be detected at only one output at a time or sent through an interferometer to demonstrate that single photons can interfere with themselves \cite{galvez2005interference,beck2012quantum}. Details about the specific experimental procedures followed by the students in this study are provided in Sec.~\ref{sec:methods_expts}.

\subsection{Resources Framework} \label{sec:bg_resources}

This study, just like a large portion of physics education research (PER), focuses on students' conceptual understanding and therefore depends on knowledge of how students learn. One theory of learning is that people learn by actively constructing knowledge themselves \cite{bada2015constructivism}. Within this constructionist view, two of the most common approaches taken in PER are identifying student difficulties \cite{heron2018identifying} and identifying the pieces of knowledge or ``resources'' students use to construct knowledge \cite{wittmann2018research}. The studies focused on discovering student difficulties can help instructors know where educational interventions may be most beneficial and aid in the development of new materials \cite{heron2018identifying, emigh2015student, wan2019probing, emigh2020research}.  
Another approach is to focus on the resources students have, so instructors can help students build off of their own ideas to learn new topics \cite{smith1994misconceptions,hammer2000student}. Instructors' can then help students identify potentially useful ideas they already have and learn how and when to apply them \cite{scherr2007modeling}.

There has been various terminology in the literature used to describe student resources or pieces of knowledge students can use to construct new knowledge (e.g., Refs.~\cite{disessa1993toward, minstrell1992facets, hammer2000student}), each with its own definition. In this work, we use the term resources, which is popular in the PER community and can encompass any size idea held by an individual student that can be used as a building block to construct additional knowledge \cite{hammer2000student,wittmann2018research}. Resources can be broad and encompass many other resources (e.g., ``coordinate systems'') \cite{sayre2008plasticity} or be more specific, around the size of a typical course learning goal (e.g, ``forces influence the motion of objects'') \cite{robertson2023identifying}. Resources are ideas that students may have learned at any point in their lives, either inside or outside of the classroom \cite{sayre2008plasticity}. They may be activated in different contexts, and do not necessarily need to be correct as long as they can be productive in at least one context, leading to the possibility of students holding seemingly contradictory ideas \cite{brown2009conceptual}. Once resources have been identified, instructors can help students refine the way different resources are activated and organized so that their ideas line up with the canonical understanding of physics \cite{robertson2023identifying, hammer2005resources}. Using the resources framework to study student learning allows us to focus our attention on the students, place value on the knowledge that they have, and understand the diverse ways different students may engage with a single context \cite{wittmann2018research}.

The resources and difficulties frameworks both have characteristics that have been identified in student reasoning, so it can be useful for instructors to consider multiple theoretical models \cite{scherr2007modeling}. Most of the work investigating student reasoning in the context of the single-photon experiments has focused on student difficulties  (e.g., Refs.~\cite{marshman2016interactive, marshman2017investigating}), so it is also important to identify resources students activate as they reason through these complex and unintuitive topics. 

\subsection{Conceptual understanding of quantum mechanics through an experimental context}

Because of the difficulty quantum mechanics poses for students, there have been many studies investigating student reasoning and conceptual understanding of quantum mechanics, with a focus on identifying specific ideas that are challenging for students \cite{singh2015review, marshman2015framework, cataloglu2002testing, carr2009graduate, krijtenburg2017insights}. Due in part to the abstract nature of the topic, new curricula that explicitly discuss quantum optics experiments have been incorporated into quantum mechanics courses \cite{baily2010teaching, kohnle2013new, malgieri2014teaching,hoehn2018students,bitzenbauer2020new}. The use of a concrete experimental context allows instructors to discuss the interpretive aspects of quantum mechanics \cite{kohnle2013new, baily2015teaching}; elicit student ideas about differences between uncertainty in quantum versus classical contexts \cite{baily2015teaching, stump2023context}; and teach about concepts, such as particle-wave duality of photons, single-photon interference, and the way quantum measurements are probabilistic \cite{kohnle2013new, malgieri2014teaching}.

Discussing the context of single-photon experiments has helped instructors and researchers identify specific student difficulties and elucidate student ideas surrounding the behavior and properties of photons. Students have been found to have difficulty reasoning about single-photon interference with a Mach-Zehnder interferometer \cite{marshman2016interactive, marshman2017investigating}. In particular, students often ignore the wave-like properties of photons, instead discussing them as point particles traversing the arms of the interferometer. Students also often do not account for interference or the relative phase shift between the two arms of the interferometer when discussing the number of photons that will be detected after the interferometer \cite{marshman2016interactive, marshman2017investigating}. Other studies have investigated nuanced student reasoning about the ontology of photons, finding that students classify photons in a variety of ways, including particle-like descriptions, wave-like descriptions, combinations of the two, or neither \cite{mashhadi1999insights,olsen2002introducing, mannila2002building, greca2003does,henriksen2018light, bitzenbauer2022toward}. These student ideas can change between contexts \cite{hoehn2018students, hoehn2019investigating} and are affected by instruction \cite{baily2010teaching}, including both the words \cite{hoehn2019investigating} and visualizations \cite{kohnle2014investigating} used by instructors.

Various kinds of classroom activities have been shown to help students improve their conceptual understanding related to the ideas of particle-wave duality of photons and single-photon interference. One of the simplest ways to have students engage with these ideas in a concrete context is to discuss what would happen in an experiment without the students interacting with an actual experimental apparatus. Classroom discussions of the single-photon experiments as thought experiments have been shown to help students distinguish between the ways uncertainty manifests in classical versus quantum models \cite{baily2015teaching}. In order to provide students opportunities to see how experimental results depend on various parameters, interactive simulations, including of a Mach-Zehnder interferometer with single photons, have been developed and incorporated into some courses. These have been shown to improve students' conceptual understanding of single photon interference \cite{malgieri2014teaching, kohnle2015enhancing} and reduce the known student difficulty of ignoring the interference of single photons \cite{marshman2017investigating, marshman2022quilts}. Videos of real data combined with diagrams of and questions about the experiment have also been shown to help students use the concept of superpositions to explain single-photon interference \cite{waitzmann2024testing}, and interactive screen experiments have led to the improvement of students' understanding of the properties and behavior of photons and the probabilistic interpretation of quantum mechanics \cite{bitzenbauer2021effect}.

To date, there have been only a few studies investigating how working with the physical apparatus of the single-photon experiments can improve student conceptual understanding of quantum mechanics. Some of the instructors who have developed and published about the use of these experiments in their courses have shown that the experiments help students be motivated to learn about the topics \cite{pearson2010hands}, self-report an improved understanding of quantum superpositions \cite{galvez2010qubit}, and correctly answer conceptual questions about entanglement and single-photon interference, such as if interference can be observed when photons in the two arms have different polarizations \cite{lukishova2022fifteen}. Our earlier work that investigates student learning outcomes in courses across many different institutions also showed that students self-report learning quantum concepts while working with these experiments \cite{borish2023seeing}. Additionally, the experience allowed many students to confirm their belief that quantum mechanics describes the physical world even as different students provided varied responses as to what is quantum about the experiments \cite{borish2023seeing}. However, all of the education research performed on students working with the physical experiments has focused on student outcomes instead of the process, thereby missing out on understanding what specific activities prompt productive student reasoning that can lead to various learning gains.

\section{Methods} \label{sec:methods}

In order to understand students' in-the-moment thinking, we performed sets of think-aloud lab sessions with students as they were working with the single-photon experiments and followed those up with individual semi-structured interviews. In this section, we first present a description of the development of the lab guides the students used, as well as a summary of the procedures the students followed while working with the three experiments during the think-aloud lab sessions. We then describe our data sources including information about the students who participated and the details of the structure of the lab sessions and interviews. Next, we explain our analysis procedure involving both content logs and interview transcripts, and end with a discussion of the limitations that occur with this kind of detailed study. 

\subsection{The experiments} \label{sec:methods_expts}

The lab guides the students worked through in our study were designed to align with the way many instructors implement the single-photon experiments in their quantum or BFY lab courses \cite{borish2023implementation}. We developed the lab guides by looking at examples in the literature and from instructors who had shared their materials with us. We then adjusted the materials to match our logistical constraints, considering the equipment we had available, the amount of time we could ask students to commit, and the way we were implementing these experiments outside of a course setting. The lab guides included metacognitive scaffolding asking the students to reflect with their lab partner on the experimental results. Prior to providing the lab guides to the students, we tested each of the lab guides twice with colleagues who had a Master's or PhD in a field of physics or astronomy outside of Atomic, Molecular, and Optical (AMO) physics. 

\begin{figure*}
    \centering
    \includegraphics[width=\linewidth]{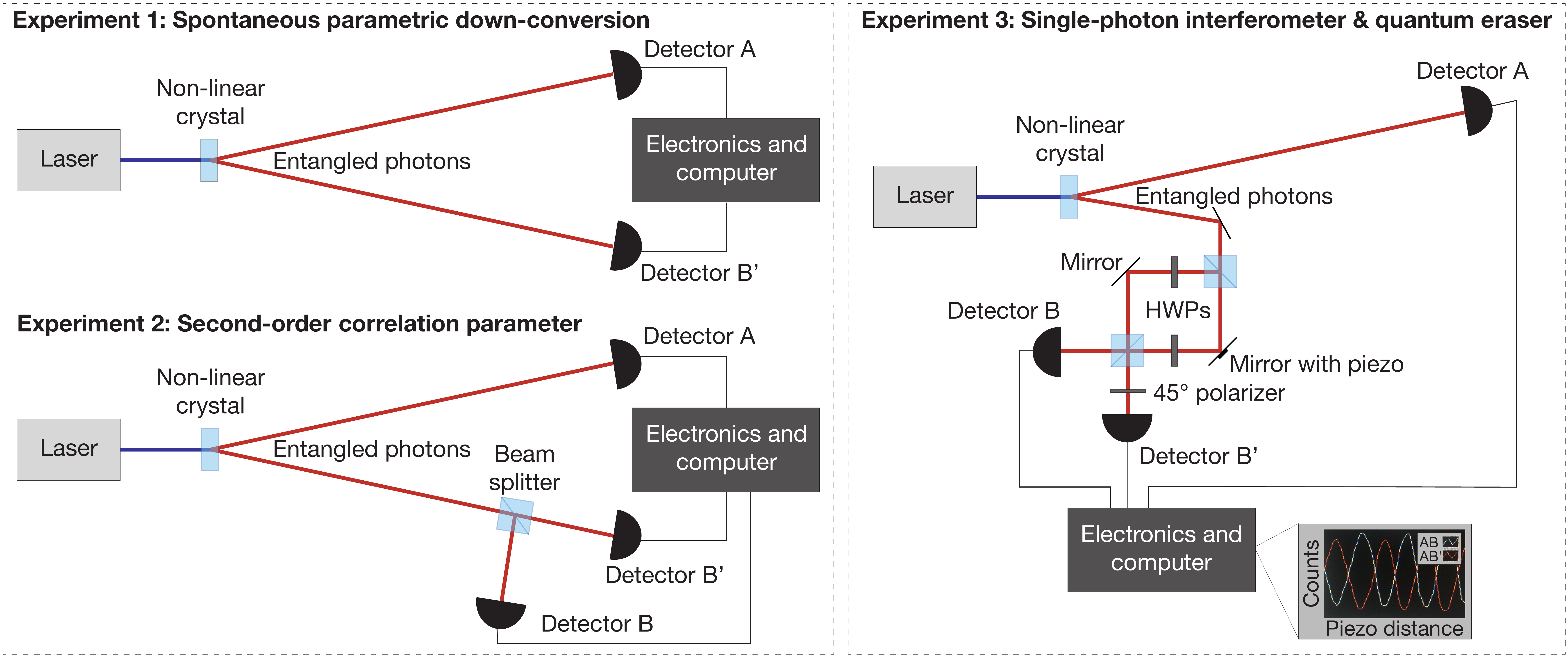}
    \caption{Schematic diagrams of the three experiments utilized in this study. In Experiment 3, the half-wave plates (HWPs) and $45^{\circ}$ polarizer were only added at the last step, for the quantum eraser.}
    \label{fig:experimentalSetup}
\end{figure*}

All three of the experiments performed by the students utilized a similar apparatus, as shown schematically in Fig.~\ref{fig:experimentalSetup}. Each began with a 405 nm laser illuminating a non-linear crystal. Inside the crystal, a small fraction of the photons from the laser were converted into pairs of spatially-entangled lower-energy photons through the process of SPDC. The two photons in each pair produced through SPDC are generated concurrently, so they would be detected at the same time in detectors equidistant from the crystal. Each experiment includes either two or three detectors, labeled A, B, and B'. The photons arriving at these detectors were converted into electronic signals, which were sent through a set of electronics that generated the number of coincidence counts for all combinations of detectors. Coincidence counts are the number of times photons arrived at each of the  detectors within the same very short time window, indicating that the detected photons were part of the same entangled pair. The single and coincidence counts were displayed on a LabVIEW computer interface. The lab guides provided to the students are detailed in Ref.~\cite{SM} and summarized below.

In our implementation of the first experiment, the students began with a setup where the laser was approximately aligned through the crystal, and one path of the down-converted photons was approximately aligned into detector A. The students had the opportunity to optimize this alignment by adjusting the tilts of the crystal and the mirrors in front of the fiber coupler attached to detector A. The students then determined where to place the fiber coupler for detector B', and placed that so as to maximize the coincidence counts between detectors A and B'. By optimizing the position of detector B', the students were able to observe spatial correlations between the down-converted photons. They additionally added a delay to the signal coming from one of the two detectors to confirm that the photons they were observing with the coincidence counts were generated at the same time.

The second experiment was a continuation of the first. When the students arrived, the laser was already aligned through the crystal, and detectors A and B' were already aligned with the pairs of entangled photons. The students began by placing a beam splitter in front of detector B' to split the path of the photons going to that detector, and then added a fiber coupler for the third detector (detector B) to detect the photons that were reflected at the inserted beam splitter. The students then measured the two- and three-way coincidence counts and used those to calculate the second-order correlation parameter, $g^{(2)}$, of the two outputs of the beam splitter. This allowed them to determine whether the light exiting the two outputs of the beam splitter was correlated (as it would be for a classical model of light, in which the amplitude of the wave splits) or anti-correlated (as it would be for single photons that can be detected at only one of the two outputs). The students obtained a value for the three-detector correlation parameter indicating that there were very few times all three detectors recorded counts at the same time, and thus that the experiment was best described by a quantum model of light. The students then made a measurement of the correlation parameter while ignoring detector A and obtained a value indicating a correlation between the outputs of the beam splitter, and thus a classical model of light. This was intended to demonstrate that there was a single-photon state only when the students accounted for the photons hitting detector A.

In between the second and third experiments, V.B. added additional optical elements to create and align a Mach-Zehnder interferometer in one of the paths of the down-converted photons. The students began the third experiment by looking at interference with a visible laser (referred to as the ``alignment laser'') that was already aligned along that same path. The students already had some familiarity with the alignment laser since they had also used it to help align the beam splitter in the second experiment. The students then spent the rest of the experiment working with the down-converted photons, with which they were immediately able to see interference. To better understand what they were seeing, the students had opportunities to play around with this setup by blocking and unblocking the two arms of the interferometer and ramping the piezoelectric actuator (piezo) that was attached to one of the mirrors in the interferometer. By using a computer-aided control system to apply different voltages to the piezo, the students were able to change the relative path length between the two arms of the interferometer by a distance on the nanometer scale, thereby causing the interference pattern to shift. To finish off the third experiment, the students placed half-wave plates in both arms of the interferometer and rotated one so that the polarizations of the photons in the two arms were orthogonal, thus removing the interference. When they then placed a polarizer (aligned at $45^\circ$ to the light in both arms of the interferometer) after the second beam splitter, the interference pattern reappeared. This is called a quantum eraser since the which-path information has been ``erased.''

\subsection{Data sources}

Our data come from four students enrolled at the University of Colorado Boulder. We recruited these students from our institution because the students needed to interact with a physical apparatus that was located there. To recruit the students, the instructor of the second semester upper-division quantum mechanics course made an in-class announcement about our study, and the students were provided a link to sign up either on their own or with a lab partner. Due to limited resources, we were able to accommodate only two pairs of students, so we selected the students who had signed up with a lab partner since we knew they would work well together. The four students who participated were juniors and seniors majoring in physics and engineering physics. By the time of the interviews, all of them had completed at least two semesters of upper-division quantum mechanics courses, a junior-level physics lab course on electronics, and at least one research experience. 

Our primary data source is think-aloud lab sessions where each pair of students worked together through a sequence of three two-hour lab sessions while being prompted to discuss their thinking out loud. During these sessions, V.B. acted as both a researcher (explaining the purpose and process of the research and prompting the students to explain their reasoning if needed) and an instructor or teaching assistant (answering questions the students had and ensuring they followed safety protocols). Since she had set-up the experimental apparatus and designed the lab guides, she was familiar with the experiments in a similar way to a typical instructor. V.B. tried to emphasize student reflection on the concepts covered in the lab and minimize prompting, so there were times when the students had long sense-making conversations on their own and other times when she asked the students questions to help them understand or explain what they were seeing, as an instructor might (see Appendix~\ref{app:allConvos} for specific examples).  
All of the lab sessions were video and audio recorded. 

After the students had completed all three experiments, they also participated in semi-structured individual interviews over Zoom, which we refer to as ``post-interviews.'' These lasted approximately one hour 
and included questions about the concepts covered in these experiments; where in the sequence of experiments the students had learned about these concepts; and other potential learning outcomes such as interest, self-efficacy, and feeling that they had seen real quantum effects \cite{borish2023seeing}. We asked demographic questions at the end of the post-interviews and found that three of the students identified as white and the fourth as white and Asian \footnote{The student specified a specific country in Asia when asked about their race, but to help preserve their anonymity, we have chosen to present only a more general racial categorization here}. Two of the students identified as women and two as men, although we use gender neutral pseudonyms and pronouns throughout the paper to help protect their anonymity. The students were compensated with a gift card for the time they spent in the lab sessions and post-interview.

\subsection{Data analysis}

Our analysis for both research questions centered around student conversations in the think-aloud lab sessions. We began our analysis by creating content logs of all the lab sessions \cite{jordan1995interaction}, in which we summarized what happened throughout the sessions and indicated particularly noteworthy moments related to our research questions. Using these content logs, we then implemented an iterative process of examining the interesting moments identified in the lab sessions and the transcripts of the post-interviews and refining our interpretations of the students' reasoning, with frequent discussions between the authors \cite{engle2014progressive}. 

Since the students spent the most time sense-making in the third experiment, we focused our analysis of resources solely on that experiment and therefore students' understanding of single-photon interference.  We began to identify student resources by watching the video recordings of the relevant conversations we had noted in the content logs. To avoid missing additional moments, we fully transcribed the third lab session for both groups. After carefully going through the transcripts and returning to watch the video clips of any moments where it seemed information could be missing (e.g., when students were using hand gestures), key conversations were chosen that demonstrated evidence of the identified resources. These were matched with student quotes from the post-interviews where the students reflected on the knowledge they had relied on while working with the experiments.

Our selection and naming of resources is based on the framework presented in Sec.~\ref{sec:bg_resources}. We endeavored to choose names for the resources that were as close as possible to the students' own words, avoiding any judgement about correctness, as the important part was finding student ideas that led to productive reasoning. Although resources can be any size, we are presenting here a set of ``small-scale'' resources that may be more in line with many PER studies since that is what instructors can most easily use \cite{wittmann2018research,sayre2008plasticity}. However, students often used many of these resources in conjunction with one another and they were not always easily distinguishable, so we also provide two larger categories of resources that we found useful when considering instructional implications.

For identifying student ideas related to what is quantum versus classical, we again began by examining relevant moments identified in the content logs. Because a difference between quantum and classical models is not specific to only one of the experiments, we chose to investigate moments from all three lab sessions. We first watched video clips of the conversations identified in the content logs, and then transcribed the relevant parts of the conversations. We additionally searched for the words ``quantum'' and ``classical'' in the automated or corrected transcripts of all three lab sessions. We performed a thematic analysis \cite{braun2006using} of these conversations where we grouped them by theme and chose the most common themes to discuss here.

\subsection{Limitations}

As with all qualitative studies with the level of detail analyzed in this study, we were only able to accommodate a small number of student participants and had to investigate one specific context. We chose to focus on four students in total divided into two groups, and these students are all enrolled at the same institution and thus have attended similar (or the same) courses. This sample is not representative of undergraduate physics students nationally, yet the ideas held by these students are likely held by other students as well.

Although we modeled our lab sessions after those used in actual courses, we investigated only one specific implementation of the single-photon experiments and it was in a non-class setting. The materials we designed were based off of materials used in other courses, but still had to be altered to fit within our specific constraints.  The context of a research setting is different than a classroom since V.B. was with the students at all times (whereas teaching assistants or instructors often rotate through groups), the students did not receive additional instruction beyond the lab sessions (for example, detailed derivations of the math demonstrated in the experiment were only available if the students decided to look up the references in their own time), and the students did not need to consider grades or lab write-ups. Additionally, it is not possible to fully separate the impact of instruction from the impact of these experiments on their own, as the lab guides and interactions with the instructor are a part of students' experiences with experiments. We therefore do not know exactly what affect V.B.'s interactions with the students had on their reasoning, although the interactions were intended to mimic that of an actual classroom environment as much as possible.

Nonetheless, there are very few studies investigating student in-the-moment reasoning while working with complex experiments such as these. Therefore, the existence of student ideas demonstrated even with a small sample and a single context can provide instructors a starting point for helping their students build off of their own ideas, while also motivating future studies in these complex lab spaces.

\section{Results: Student resources for understanding single-photon interference} \label{sec:resources}

To answer our first research question, we identify ideas students employ in the third lab session (the single-photon interferometer experiment). There, the students were compelled to make sense of the experimental data that simultaneously demonstrated particle-like and wave-like properties of single photons. This sense-making process consisted of the students discussing their observations with their lab partners and collectively building on various resources the students came in with until they had generated an explanation with which they were satisfied. Using the transcripts of the third think-aloud lab sessions, as well as the post-interviews, we identified resources the students utilized and list them in Table~\ref{tab:resourceComponents}. We group some of the resources into broader categories because students often used several resources in the same category in similar ways in a given moment.

\begin{table*}[htbp]
\centering
\caption{Identified resources and broader categories we assigned them (when applicable). The resource \emph{Thinking in terms of information} refers to the way information (in this case, which-path information) can be used as a quantifiable object to understand a physical system.}
\begin{tabular}{>{\hangindent=1.5em}>{\raggedright}p{9cm} p{6cm}} 
    \hline \hline
    \textbf{Resource} & \textbf{Category} \\
    \hline 
         Waves can constructively and destructively interfere & Knowledge of classical wave interference \\
         Need two things for interference to occur & Knowledge of classical wave interference \\ 
         Things at same place and time can interfere & Knowledge of classical wave interference \\ 
         Orthogonal things do not interfere & Knowledge of classical wave interference \\
         Waves can split and recombine & Knowledge of classical wave interference \\
         Knowledge of what laser interference looks like & Knowledge of classical wave interference \\
         Wave functions represent probability distributions & Knowledge of wave functions \\
         Wave functions can constructively and destructively interfere & Knowledge of wave functions \\
         Wave functions have phases & Knowledge of wave functions \\
         Wave functions have spatial components & Knowledge of wave functions \\
         Wave functions have temporal components & Knowledge of wave functions \\
         Quantum outcomes are probabilistic & -\\
         Thinking in terms of information & - \\
         \hline \hline
\end{tabular}
\label{tab:resourceComponents}
\end{table*}


We begin this section by presenting summaries of the two groups' progress through this lab session, including specific moments of sense-making where these resources were identified. These conversations demonstrate what resources the students activated and how the resources were productively used in context. We then synthesize these results by comparing the resources activated by the two groups, discussing where the students may have acquired these resources, and suggesting implications for instruction. 

\subsection{Anwar and Ori}

Anwar and Ori began the single-photon interferometer experiment by discussing what would happen to both single photons and a laser when passing through the Mach-Zehnder interferometer. The students predicted that single photons would not interfere with themselves because photons pick a path and go one way or the other each time they encounter a beam splitter. Before verifying or falsifying this prediction, they sent the alignment laser through the interferometer, seeing it splitting, recombining, and interfering, as they had expected.

Anwar and Ori then moved on to working with the heralded single-photon source where they noticed that the observed coincidence counts were changing rapidly. When asked by V.B. how the fluctuations compared with fluctuations they had seen in the previous experiments, they responded that this was different than before, but did not try to understand why. The students then changed the voltage sent to the piezo and noticed that the counts in the two outputs of the final beam splitter of the interferometer were changing as well. Their ensuing conversation, after they were prompted to explain what they were observing, is shown in Appendix~\ref{app:AnwarOri1}.

There, Anwar and Ori pointed out that the pattern of coincidence counts increasing and then decreasing was similar to what they had seen with the alignment laser (lines 7-10 and 14). They were activating the resource \emph{Knowledge of what laser interference looks like} since they referenced the visual appearance of varying intensities of the laser after it had passed through the interferometer. This resource, however, may also have contributed to the way the students did not initially realize that the experimental results they were observing did not line up with their prediction that single photons would not interfere. The students were still thinking about the alignment laser even as they began working with single photons and therefore did not realize their prediction did not match the experimental results they were seeing until it was pointed out by V.B. (lines 12-20).

When prompted to reason through the experimental evidence of single-photon interference, Anwar and Ori used various resources in the categories Knowledge of classical wave interference and Knowledge of wave functions. In response to V.B.'s question about what is a requirement for interference, Anwar and Ori activated the resources \emph{Need two things for interference to occur} and \emph{Waves can constructively and destructively interfere} (lines 25-35). When further prompted to think about how they could relate this to the experiment, the students brought up wave functions for the first time (lines 40-44), discussing how a photon's wave function represents the different probabilities of it traversing each path of the interferometer (the resource \textit{Wave functions represent probability distributions}). The students then went on to explain this by thinking about wave functions mathematically, activating the resources \textit{Wave functions have spatial components} and \textit{Wave functions have temporal components} (lines 55-60) as well as \textit{Wave functions have phases} (lines 61-63). Putting all of this together along with the idea that wave functions can constructively and destructively interfere (lines 64-83), Anwar concluded: ``So it doesn't make immediate sense, but wave functions. Wild.''

Anwar and Ori continued to use similar reasoning as they progressed to the next task in the lab guide: blocking one arm of the interferometer when interference was at a minimum for one set of coincidence counts. To explain what they were seeing, they again used resources related to wave functions, including \emph{Wave functions have phases} (lines 6 and 22 in Appendix~\ref{app:AnwarOri2}), \emph{Wave functions represent probability distributions} (lines 12-15, 24, and 48-75), and \emph{Wave functions can constructively and destructively interfere} (lines 21-38, and 48-68). They discussed how, when they were blocking one arm of the interferometer, there was one wave function instead of two, so there was no phase. This removed the interference pattern and thus affected the probabilities they were seeing as counts on the detectors. By the end of this conversation, both students were content to use the idea of wave functions to understand the experimental results they were observing. 

Next, Anwar and Ori moved on to looking at the second-order correlation parameter, which provided evidence that the photons were exhibiting particle-like behavior at the same time they were also interfering. Immediately after reading off the second-order correlation parameter from the computer interface, the students again used their wave function resources, in particular \emph{Wave functions represent probability distributions} (lines 14-23 in Appendix~\ref{app:AnwarOri3}), to reason through why these two types of behavior can exist at the same time. Ori concluded ``the wave function is acting like a wave, but it's really just determining the probability of the particle going into one or the other.'' Anwar also mentioned how they were thinking about the spatial dependence of wave functions (lines 32-37). 

The last part of the experiment involved the students implementing a quantum eraser. After putting half-wave plates in both arms of the interferometer and rotating one of them so the two paths had orthogonal polarizations, Anwar and Ori noticed that the interference had been eliminated. They used their knowledge of classical wave interference, in particular the resource \emph{Orthogonal things do not interfere}, to understand what they were seeing. The students gestured with their arms to represent the orthogonal polarizations while discussing this with each other. Ori later explained ``they're just not interfering... because, like, the waves are perpendicular to each other.'' 


Anwar and Ori continued to use the resource \emph{Orthogonal thing do not interfere} to understand the experimental results after placing a polarizer oriented at $45^{\circ}$ in front of detector B', but not detector B. The students noted that the behavior of the counts in the two detectors differed. Ori described what was happening: ``So AB' is changing a lot. And AB is not. Which suggests that interference is having a big effect going into B' but not B.'' Anwar made sense of this by saying,
\begin{quote}
    I think that that makes sense, right? Because... 
    the only light that's allowed to hit the B' detector has the same polarization. Right? So the effect of the construction or the deconstruction will actually matter... 
    Versus the light that goes into B, like you know, the polarizations might be orthogonal, so messing with the beam length or messing with the beam path length doesn't necessarily mean that they're gonna construct or deconstruct.
\end{quote}

Anwar and Ori used this same resource once again to predict and explain what would happen when they put another $45^{\circ}$ polarizer in the setup, this time in front of detector B. Before looking at the experimental results, Ori predicted,
\begin{quote}
    We're doing the exact same thing that we just did, so it should behave the same as the other one. Because they're going to be orthogonal, but then they're going to collapse down to the same thing. And then like, is that when the interference happens?
\end{quote}
After seeing that the interference did indeed reappear in detector B after the insertion of the second polarizer, Ori explained, 
\begin{quote}
    It makes sense. I mean, symmetrically like it makes sense. But it also makes sense because-- I guess they were orthogonal. When they're coming back together, at some point, they're gonna be at the same place at the same time and interference is happening. So it's like the interference pattern is reintroduced by the polarizer. That makes sense. It's weird, but cool.
\end{quote}
Here, they also used the resource \emph{Things at same place and time can interfere}. Both Anwar and Ori ended the lab session satisfied that they had been able to explain the experimental results they saw.


\subsection{Kiran and Luce}

Kiran and Luce gave a less definitive answer at the start of the single-photon interferometer experiment about whether they thought single photons would interfere. When asked what they would expect to see at the outputs of the interferometer when they sent in single photons, Luce discussed how they were ``not sure if [they] would see any difference'' since the photons would take one path or the other and therefore end up at only one of the detectors. Kiran followed this up by distinguishing between measurements of individual photons and a set of many photons:
\begin{quote}
    Surely, we can only detect one at a time. If you do many single photons... as in a laser, for instance, I presume that the interference will be controlling which detector they preferentially go to.
\end{quote}
It is not clear whether they were differentiating between what they would expect with single photons compared with a laser or compared with many single-photon states at the same time (which is different than the coherent state coming from a laser, although that distinction may be beyond the students' knowledge).

Kiran and Luce saw fluctuating counts that were indicative of interference with both the alignment laser and single photons. They began the experiment looking at interference with the alignment laser and expressed that it was behaving as they expected when they tapped the table and varied the voltage sent to the piezo. They then switched to using the heralded single photons and noticed an immediate imbalance between the counts in the two detectors, but were not certain if that was strange or expected. As they changed the piezo settings, Luce noted that the piezo voltage was affecting the relative counts on the two detectors:
\begin{quote}
    And they're oscillating. Whoa. Whoa! Whoa! Oh, that's cool. Okay, it's not as cool anymore. It is kind of still kind of cool. Anyways, I changed the volts [being sent to the piezo] because I thought maybe that somehow the voltage zeroed on the piezo was favoring one [detector] over the other. And by shifting a few volts, they are now similar. So perhaps my hypothesis is confirmed... Zero [volts] prefers B'. So two-ish [volts], maybe, it's making [the counts in both detectors] even, but they're oscillating so much. So maybe now there's interference? I don't know.
\end{quote}
Kiran later pointed out that the counts from the two detectors were anti-correlated with each other: one increased as the other decreased. To explain this, they said, ``I kind of just want to say statistics and leave it at that.'' The students did not provide an additional explanation, so we were not able to classify their reasoning in this section of the lab session into specific resources.

The first clearly identifiable resources Kiran and Luce activated were in the next part of the experiment where they recorded numbers of counts when blocking and unblocking the two arms of the interferometer. In response to V.B.'s question about what effect could be seen in the detectors when the piezo position was adjusted while one arm of the interferometer was blocked, Luce said,
\begin{quote}
     Interference! We're changing the length of the interferometer, so if it's a wave, there's going to be some constructive or destructive interference between the two waves when they recombine, which is going to affect our measurements at either of the detectors. And that's all I feel comfortable stating.
\end{quote}
Kiran then tied this idea in with the question asked by stating, ``Of course there's no recombination if we block one of the arms.'' Although Kiran and Luce did not articulate a full explanation for single-photon interference at this point, they were activating the resources \emph{Waves can constructively and destructively interfere} and \emph{Waves can split and recombine} to understand the changes in the counts they were seeing. They continued to investigate the effects of various actions on the coincidence counts by slowly varying the piezo voltage and blocking one of the arms of the interferometer when at an interference minimum.

Another interesting sense-making moment was when Kiran and Luce looked at the second-order correlation parameter while also seeing interference (see Appendix~\ref{app:KiranLuce1}). There, the students again used the resources \emph{Waves can constructively and destructively interfere} and \emph{Waves can split and recombine}, while also activating the resources \emph{Knowledge of what laser interference looks like} and \emph{Quantum outcomes are probabilistic}. When asked what was happening at the first beam splitter, Luce pointed to different parts of apparatus, referring to them as either ``laser'' or ``photons.'' They were distinguishing between the parts of the apparatus where they were thinking of the photons as having wave-like properties (in the interferometer) and where they were thinking of the photons as having particle-like properties (at the detectors). After acknowledging that there was no laser in the interferometer, yet they were still seeing interference, Luce said, ``So somehow the waves are going from waves to single photons here.'' To which Kiran responded, ``Or could it be that the single photons were waves all along?'' Kiran followed this up by discussing probabilities indirectly (``what might happen'') as a way to avoid thinking about splitting individual photons. 

Kiran and Luce then varied the setting of the piezo while still looking at the second-order correlation parameter. Kiran summarized this as ``And what we're doing is sweeping through interference preferring one arm to the other to even. And yet, all of the time, they are single photons doing the interference.'' To which Luce responded: ``Perhaps, it was a wave all along. It sounds like an Agatha Christie story.'' By this point the students had convinced themselves that the single photons did have some wave-like properties, and the researcher summarized what they had been saying adding in the terminology of superpositions.

To finish off the lab, Kiran and Luce implemented a quantum eraser. While putting half-wave plates into both arms of the interferometer and rotating one so the polarization was rotated by 90 degrees, Luce said,
\begin{quote}
    And there would no longer be interference because they're polarized differently. They're ninety degrees to each other. They're orthogonal!
\end{quote}
They were activating the resource \emph{Orthogonal things do not interfere}. While discussing how polarizers work, Kiran also activated the resource \emph{Quantum outcomes are probabilistic}:
\begin{quote}
    I think classically [the polarizer] filters out such that you get like the reduced dot product or whatever. And quantumly, you just allow them probabilistically to the same average.
\end{quote}
Although this is not directly connected to the students' reasoning about single-photon interference, it shows an example of activating the same resources in another context, in this case to understand how part of the apparatus worked.

The final resource, \emph{Thinking in terms of information}, was utilized by Kiran and Luce at the end of the experiment to explain the idea of the quantum eraser. V.B. helped guide the students through understanding the role of the polarizer after the interferometer by explaining that when the photons in the two arms of the interferometer were polarized orthogonally to each other, an experimenter could measure which arm of the interferometer a photon went through, and therefore the interference was eliminated. When asked what the polarizer was doing, Kiran first brought up the term ``information'' (line 4 in Appendix~\ref{app:KiranLuce2}). They then went on to explain how this idea could be used to describe the reappearance of interference:
\begin{quote}
    ...if you receive it and you see its polarization, and you know how it started, you know which waveplate rotated it. But if there's the polarizer in the way, just information-wise, you don't--- everything you receive is the same polarization and it could have come from one or the other, or classically the wave... And therefore, since it could have come from either, it can do the interference.
\end{quote}
This led the students to ultimately conclude that the polarizer was erasing the which-path information, leading to the reemergence of the interference pattern.

\subsection{Discussion of student resources}

\begin{figure*}
    \centering
    \includegraphics[width=1\linewidth]{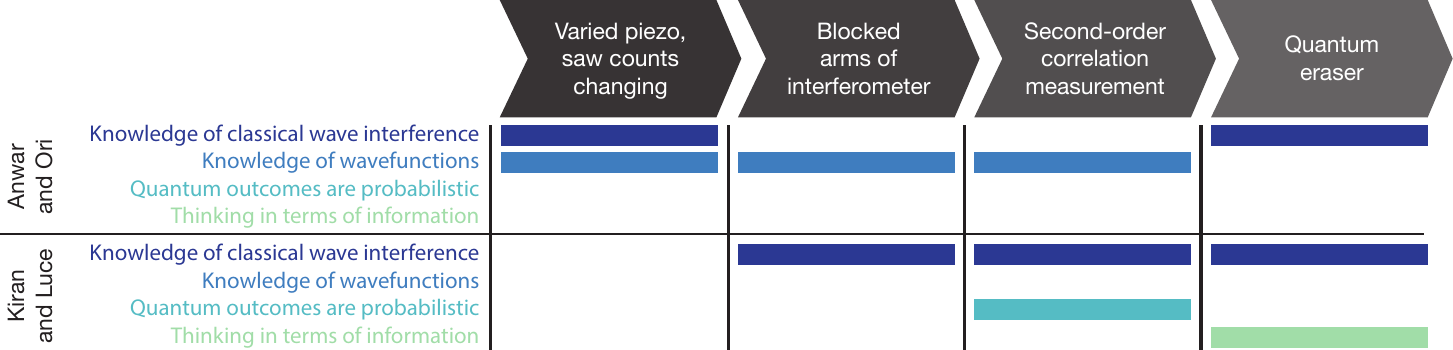}
    \caption{Categories of resources (or resources for the two that did not fit into categories) activated by each pair of students to make sense of single-photon interference during different parts of the single-photon interference experiment.}
    \label{fig:resources}
\end{figure*}

By the end of the experiment, both groups of students were ultimately able to make sense of the fact that they were seeing both particle-like and wave-like behavior of photons at the same time. However, they activated different, yet overlapping, resources to do so. Both groups used their prior knowledge about classical wave interference when first reasoning through the existence of interference and later explaining the quantum eraser.  Anwar and Ori spent a large portion of their lab session discussing wave functions, activating both conceptual and mathematical resources including the idea that wave functions represent probabilities of the paths the photons may traverse (the resource\emph{Wave functions represent probability distributions}). Kiran and Luce, on the other hand, never explicitly mentioned wave functions, but instead activated the related resource \emph{Quantum outcomes are probabilistic}. They additionally used the resource \emph{Thinking in terms of information} when explaining the quantum eraser. The broad categories of resources (or resources for the two that did not fit into categories) that both groups of students activated at different points in the experiment are shown schematically in Fig.~\ref{fig:resources} to provide a visual overview of the two groups' reasoning. The specific resources activated by each pair of students are indicated in Table~\ref{tab:specificResourcesDuringExpt}. 

\begin{table*}[htbp]
    \centering
    \caption{Resources identified as being activated by the students to reason through single-photon interference at different times during the experiment. It is possible the students used these resources at more times than indicated, but we were only able to identify resources when the students clearly verbalized their reasoning.}
    \begin{tabular}{p{0.25cm}lcccc}
    \hline \hline
         & &  \parbox{2cm}{Varied piezo,\\ saw counts changing} & \parbox{2cm}{Blocked\\arms of\\interferometer} & \parbox{2cm}{Second-order\\correlation measurement} & \parbox{2cm}{Quantum\\eraser}\\
         \hline 
         \multicolumn{2}{l}{Anwar and Ori}  &  &  &  & \\
         &Waves can constructively and destructively interfere & \checkmark &  &  & \\
         &Need two things for interference to occur& \checkmark &  &  & \\
         &Things at same place and time can interfere & &  &  & \checkmark \\
         &Orthogonal things do not interfere &  &  &  & \checkmark \\
         &Waves can split and recombine &  &  &  & \\
         &Knowledge of what laser interference looks like & \checkmark &  &  & \\
        &Wave functions represent probability distributions& \checkmark & \checkmark & \checkmark & \\
        &Wave functions can constructively and destructively interfere & \checkmark & \checkmark &  & \\
        &Wave functions have phases & \checkmark & \checkmark &  & \\
        &Wave functions have spatial components & \checkmark &  & \checkmark & \\
        &Wave functions have temporal components & \checkmark &  &  & \\
        &Quantum outcomes are probabilistic &  &  &  & \\
        &Thinking in terms of information&  &  &  & \\
         \hline
         \multicolumn{2}{l}{Kiran and Luce}  &  &  &  & \\
         &Waves can constructively and destructively interfere &  & \checkmark & \checkmark & \\
         &Need two things for interference to occur&  &  &  & \\
         &Things at same place and time can interfere & &  &  & \\
         &Orthogonal things do not interfere &  &  &  & \checkmark\\
         &Waves can split and recombine &  & \checkmark & \checkmark & \\
         &Knowledge of what laser interference looks like &  &  & \checkmark & \\
        &Wave functions represent probability distributions&  &  &  & \\
        &Wave functions can constructively and destructively interfere &  &  &  & \\
        &Wave functions have phases &  &  &  & \\
        &Wave functions have spatial components &  &  &  & \\
        &Wave functions have temporal components &  &  &  & \\
        &Quantum outcomes are probabilistic &  &  & \checkmark & \\
        &Thinking in terms of information&  &  &  & \checkmark \\
         \hline \hline
    \end{tabular}
    \label{tab:specificResourcesDuringExpt}
\end{table*}

The resources in the category Knowledge of classical wave interference were used by both groups of students at various stages of the experiment. Kiran and Luce used these resources throughout, whereas Anwar and Ori used them at the start and end, with discussions of wave functions in the middle. Both groups used the resources \emph{Waves can constructively and destructively interfere} and \emph{Knowledge of what laser interference looks like} to understand that the single photons were interfering, and they both also used the resource \emph{Orthogonal things do not interfere} while investigating the quantum eraser.  The students had acquired most of these ideas in previous courses, such as electricity and magnetism. The students may have additionally gained the resources related to laser interference and the splitting and recombining of light from previous experiences with different types of interferometers in lab classes or research experiences. This category of resources helped students recognize constructive and destructive interference and realize that in order for single photons to interfere, they needed to somehow be divided and then be the same when coming back together. 

Students who were less familiar with interferometers talked about how working with the alignment laser at the start of the third experiment was particularly useful. When asked in the post-interview what parts of the experiment helped them learn about particle-wave duality, Luce said,
\begin{quote}
    ...setting up the alignment laser and seeing the path that the particle, or like the laser, would take. And seeing that it could actually go two ways and then recombine. And then turning on the actual [405 nm] laser and seeing that data... I think if I had just jumped in to just kind of play with the normal laser, with an already aligned optic table, I wouldn't have actually internalized that. But having  seen--- set it up myself and seen that what I internalized as like wave behavior... was also observed on something that should just be a single particle really helped me understand that there's a dual nature.
\end{quote}
It was important for them to understand wave interference in an easy to visualize way to be able to translate that idea to single photons. Knowledge about classical light interference has been identified as a resource in the context of simulations of the single-photon experiments as well \cite{marshman2022quilts}.

Resources in the category Knowledge of wave functions, which  encompasses ideas related to both conceptual and mathematical understanding of wave functions, were also activated frequently by Anwar and Ori. They activated the resources \emph{Wave functions represent probability distributions}, \emph{Wave functions can constructively and destructively interfere}, \emph{Wave functions have phases}, \emph{Wave functions have spatial components} and \emph{Wave functions have temporal components}. These resources were used to reason that by changing the path length of one arm of the interferometer, the students were adding a relative phase between two parts of the wave function (although they referred to this as two separate wave functions). This was enabled by the fact that wave functions can change with time and depend on spatial position. The phase affected the probability of the photons going into the two detectors since wave functions can constructively and destructively interfere.

Both students who used the resources in the category Knowledge of wave functions discussed how they had acquired this resource in their quantum mechanics courses. When asked in the post-interview which specific parts of their quantum mechanics courses were most relevant when working through these experiments, Ori said,
\begin{quote}
    I think having an understanding of wave functions was helpful, especially on the third day, like mathematically, and understanding that there's a time and spatial component, and that you can have a phase shift. That helped things click too.
\end{quote}
Anwar also discussed in their post-interview how a good conceptual understanding of wave function interference had helped prepare them for this lab session: 
\begin{quote}
    Honestly, I feel like a lot of Quantum 2, at least the way that [my instructor] taught it, had a lot of really great visuals for how wave functions construct and deconstruct... [My instructor] prepared us really well to take exam questions that asked qualitatively: Would this be higher or lower? Would this be positive or negative?... That was a super valuable part of my quantum physics education in general, I think. Like having that, that kind of is like something I can intuitively think about, where stuff constructs and deconstructs, how that might evolve with time.
\end{quote}

Although Kiran and Luce did not explicitly mention wave functions while working through the third experiment, they used the resource \emph{Quantum outcomes are probabilistic} in a similar way. This resource was used to help them understand that even though a single photon cannot be thought of as splitting in half at a beam splitter, the probability of the photon going the different ways can be split. This resource may be a less developed version of the resource \emph{Wave functions represent probability distributions}. However, it is not clear why Kiran and Luce did not discuss wave functions while working with the experiment, since both of them did mention wave functions when explaining concepts from the experiment in their post-interviews.

The last resource, \emph{Thinking in terms of information} was used by only Kiran and Luce, and it was an important part of their sense-making about the quantum eraser. In the post-interview, Kiran described this resource as ``contextualizing the eraser as a matter of information'' since they were ``prepared to think about information... as an actual quantifiable thing.'' They used this resource to explain how they only saw interference when there was no available information about which path of the interferometer the photons took. This resource may also be a replacement for some of the resources related to wave functions, since in the post-interview Kiran discussed the connection between the concept of information and quantum states:
\begin{quote}
    ...I should have been confused classically, because it's confusing, but having done it, I was not very confused... because I was thinking about, in terms of information. And the whole time I was thinking, having taken the Quantum Computing class, I was thinking about just a product of kets, and if I have measured the information ket, then everything collapses. But if I have separated the path information from the polarization information then that makes it make sense, I suppose. So that last little twist was the, it was the proof of what was going on.
\end{quote}
In addition to having taken two upper-division quantum mechanics courses, Kiran had also taken a quantum computing class and had watched various videos related to information on their own. This may explain why they were able to activate this resource even though it is not a common resource for undergraduate physics students \cite{marshman2017investigating}.

The students acquired all of the resources we identified, at least to some degree, from their coursework. Some of the resources were learned in quantum mechanics courses (both traditional quantum mechanics courses, as well as quantum computing courses), and others came from courses in other fields of classical physics or mathematics. The students may have also added on to some of these resources by watching videos online or participating in research experiences.

\subsection{Instructional implications related to student resources}

We envision instructors using these results in two ways. First, knowing what resources students may activate can allow instructors to incorporate the single-photon interferometer experiment at an appropriate time in the curriculum. Second, seeing the context in which the students activated different resources may allow instructors to better aid their students in constructively building off of their own ideas to understand particle-wave duality of photons. This may also help instructors understand how experiences with a physical experiment can facilitate this learning process. Keeping in mind that there are likely many additional resources other students would use in a similar context, we present some implications for instruction based on the students in our study.

\emph{Instructors could consider utilizing the single-photon interferometer experiment after students have learned about classical wave interference and how quantum states are connected to probabilities of outcomes.} The focus of the single-photon interferometer is not on classical wave interference, yet understanding that two waves can constructively and destructively interfere and what that looks like may be necessary for students to understand that they are seeing interference with single photons. Additionally, it may be helpful for students to have at least some idea that measurement outcomes in quantum mechanics are probabilistic. Both sets of students in our study activated resources related to these ideas in order to understand their experimental results, so instructors may want to ensure their students have learned these topics before working with the experiment. 

The decision of when to incorporate an experiment into a course is important as it affects what conclusions the students can come to while working with the experiment and what learning they may be primed for afterward. From our data, we cannot definitively say which concepts the students should know ahead of time versus which ones they could best learn in the context of these experimental observations, but students may engage with the ideas more productively when the experiments occur after or in conjunction with units on topics such as wave functions, superpositions, and measurement probabilities (if included in a quantum mechanics course). For instructors implementing this experiment in a BFY lab course (or any other context where the students may not have previously taken a quantum mechanics course), they could assign activities that help students develop these resources alongside the experiment.

\emph{Instructors could consider providing students experience with a visible laser interferometer before working with the single-photon interferometer.} This could be especially helpful for students who have not seen one before or who would benefit from a reminder about classical wave interference. A visible alignment laser may already be part of the experimental setup, so it could be easy to incorporate this into the students' lab experience (as we, and many others, do) if time allows. A warm-up about classical light interference has been shown to help students working through activities with a simulation of single photons in a Mach-Zehnder interferometer \cite{marshman2022quilts}, and this may extend to students working with the physical apparatus as well. Alternatively, the students may have other resources related to interference that instructors could help them activate, such as by discussing other interferometers they have interacted with in prior research or coursework.

\emph{Even for potentially unintuitive topics like particle-wave duality, instructors could help students activate their own resources, whether or not they are fully-developed and canonically correct.} The students we studied did have the resources they needed to be able to make sense of experimental evidence of single-photon interference, even though they did not necessarily phrase their ideas in the same way their instructors might. For example, neither group initially used the word ``superposition'' and only one emphasized ``wave functions,'' but both groups still discussed how the photons had some probability of exiting the beam splitter in each of the two directions. Connections between the specific resources activated in the different parts of the experiment (see Table~\ref{tab:specificResourcesDuringExpt}) may also help instructors decide which activities to include as part of their lab guides and when they may be able to help students activate and expand on their existing resources. 

\section{Results: Student ideas of what is quantum versus classical} \label{sec:quantumVsClassical}

We additionally investigated student ideas related to what is quantum versus classical about the three experiments, another type of student reasoning about which instructors care. Not only is understanding the differences between quantum and classical models of light a common learning goal of instructors using these experiments \cite{borish2023implementation}, but it is also important to understand what students think is quantum about these experiments since students have been shown to obtain benefits from seeing quantum effects themselves \cite{borish2023seeing}. In this section, we describe the three most prevalent themes we identified in the sequence of think-aloud lab sessions:
\begin{itemize}
    \itemsep0em 
    \item Waves and interference are classical
    \item Quantum mechanics is math
    \item Quantum can be turned off
\end{itemize}

Although each of these themes showed up at various times throughout the second and third lab sessions, there were indications that some students' ideas related to the themes \emph{Waves and interference are classical} and \emph{Quantum mechanics is math} may have changed during this time as well. In the following subsections, we present student quotes that exemplify these themes with an emphasis on the way these ideas may have changed due to the students' experiences working with the experiments. This demonstrates how experiments may be able to help students understand some nuances of what it means for something to be considered quantum.

\subsection{Waves and interference are classical}

The first theme we identified encompasses two primary ideas: (1) waves and interference are classical concepts and (2) quantum objects cannot be split. Although these may seem like different ideas, they were often used in conjunction with each other since one line of student reasoning was that waves can split and splitting is a necessary precursor to recombining and therefore interfering. Both of these ideas are similar to already demonstrated student difficulties of ignoring both the wave nature of photons and the interference of single photons \cite{marshman2017investigating}.

By the end of the second lab session, students in both groups associated waves and interference with classical behavior. While summarizing the second experiment at the start of the third lab session, V.B. prompted Anwar and Ori to think about what happens to a single photon at a beam splitter. Anwar explained that the photon would go one way or the other and asked if that was classical, to which Ori responded,
\begin{quote}
    No, because classical would be if it was like a wave-y boy that got split somehow, but quantum mechanically it's a quantum thing that you can't split up.
\end{quote}
Ori was saying how waves (and objects that can somehow be split) are classical whereas single quanta cannot be split. 

The other group also discussed this idea while summarizing the second experiment at the start of the third lab session. While describing what they had seen, Luce said,
\begin{quote}
    We measured [the second-order correlation parameter] at one which meant it was classical. Split like a classical wave. So, when we have [detector] A on, it's like a particle going to one or the other. And then when we turn off [detector] A, it's like a wave.  
\end{quote}
They were associating waves with being classical and particles with being quantum. It is not clear if this is an idea the students came in with or if it came about, at least in part, due to the way the lab guide described how waves' intensity is split in the classical model of light, but that the probability of the path the photon takes is split in the quantum model. 

This idea that interference is classical probably contributed to some of the students predicting that single photons would not interfere. When predicting what would happen at the start of the third lab session, Anwar and Ori discussed how at both beam splitters in the interferometer each single photon would ``choose a path,'' so there would be no way for a photon to recombine with itself since it could not split up in the first place. They therefore did not expect to see an interference pattern, so they were then surprised when they did see evidence of interference with the single photons. Ori even asked, ``So why is it behaving classically?'' indicating that they were still thinking of interference as a classical phenomenon.

This idea of waves being classical continued through the end of the third experiment for some students, but it also started to change for others. When measuring the second-order correlation parameter while seeing interference during the third experiment, Ori said,
\begin{quote}
    And so we're seeing that it's really close to zero, which means that it's acting quantum mechanically. Oh. So we're saying it's acting quantum mechanically when it's acting like a wave because it's a wave function. Like, there's interference.
\end{quote}
They started to understand that wave behavior and interference are also a part of the quantum description of light.

\subsection{Quantum mechanics is math}

Throughout the majority of the sequence of experiments, the students thought of quantum mechanics as very mathematical. This is not surprising given that their quantum mechanics courses did not include lab components and that this is a common view of many students \cite{johnston1998student, singh2015review}. Both groups of students first discussed this idea near the end of the second experiment, when they measured a classical value for the non-heralded second-order correlation parameter.

When discussing the difference between calculating the second-order correlation parameter using counts from two detectors (un-heralded) versus three detectors (heralded), Kiran and Luce brought up the separation between the mathematics and the experimental setup:

\noindent \hangindent=.5cm \textbf{Luce}: We need to have a third detector to like have the entanglement work. Because... that place here, it gets split into the two, and then when we ignore one of the two, this one gets wonky.

\noindent \hangindent=.5cm \textbf{Kiran}: We're not, like, not measuring. It's still interacting. We're just not including it in the computation.

\noindent \hangindent=.5cm \textbf{Luce}: We're not including it relative to this stuff though.

\noindent \hangindent=.5cm \textbf{Kiran}: Numerically, we're not.

\noindent \hangindent=.5cm\textbf{Luce}: Numerically, we're not. It's still being measured, but like in our math, it's not being accounted for. And that's all quantum mechanics is is math.

\noindent They are pointing out that even though photons are still hitting detector A, those counts are not being incorporated into the students' calculation when they obtain a value that indicates classical behavior. This emphasizes for the students the idea that quantum mechanics is just math.

Anwar and Ori also discussed how quantum mechanics makes more sense mathematically. After V.B. pointed out to these students that they did not have single-photon states when they stopped accounting for detector A, Anwar said, ``It is kind of problematic to think of them as photons, right. Because they're states, which makes more sense on paper.''  However, by the time Anwar and Ori got to the end of the third lab session, they discussed how they had seen experimental evidence for some of the math. When asked if they wanted to discuss anything else, they began to realize that the math they had been using was motivated by experiments:

\noindent \hangindent=.5cm \textbf{Ori}: Yeah, it's weird. And it's making me like not ignore things that I ignore to understand the math, which is weird. But---

\noindent \hangindent=.5cm  \textbf{V.B.}: Can you give an example?

\noindent \hangindent=.5cm \textbf{Ori}: I guess, like, you're told that there's like measurements affect the outcome. But then I was like, I mean, you're just saying that so like, I'll treat it that way because that's what you've said it was and you taught me that I was supposed to treat it a certain way after it's been measured. But like, I mean, I still don't understand the mechanics of how or why the measurement affects it. And I guess that's probably a big area of research happening, like nobody really knows what's happening there. But it's interesting to see it in action. And see that the wave--- like, I can see why someone would think that a wave function exists after this. Like, how did someone come up with that? And why would they think it makes sense? Interference is a strong reason why.

\noindent \hangindent=.5cm \textbf{Anwar}: That's fair. Yeah. This is good evidence for how, even though it's like not really logical that there is a wave function, at least logical in the sense of our own day-to-day interactions with the world. But this is good evidence for it. It's fun to observe it.

Ori expanded on this idea in their post-interview. When asked what in particular helped them learn some of the quantum concepts, they talked about seeing the physical apparatus:
\begin{quote}
    The experiment that we did in lab 3 where we split up the photons and then recombined and were looking at interference. I never would have--- I guess I could have maybe mathematically shown that that would happen. But it's--- that was enlightening to see it actually happening and seeing the evidence of it happening.
\end{quote}
By seeing experimental results that can only be explained using the mathematics of quantum mechanics, the students began to observe that quantum mechanics truly describes the world. 

This is similar to other work showing that students improved their belief that quantum mechanics describes the physical world by seeing quantum effects in experiments \cite{borish2023seeing}. While it is important for students to understand and believe that quantum mechanics describes the world, in the end, just like the rest of physics, it is only a mathematical model. Nonetheless, helping students understand the nuances of that, and how good of a model quantum mechanics is, may be valuable.

\subsection{Quantum can be turned off}

The third theme, which may be more linguistic in nature, consists of how students talked about quantum mechanics as a way particles can act or something that can be turned on and off, indicating that something could be quantum in one context, but classical in another. This idea showed up in both the second and third lab sessions when the students were measuring the second-order correlation parameter. 

Students' first indicated how an object could behave quantum mechanically in one context but not in another when discussing the two-detector second-order correlation parameter near the end of the second lab session. When trying to understand the difference between the two- and three-detector measurements, Luce said, ``So by ignoring the third detector, no longer measuring photons in that detector, we are turning off quantum mechanics.'' They said something similar when summarizing this second experiment at the start of the third lab session as well: ``...when we stopped taking data over there, things got a little wild and weird... classical even.'' They discussed how the results could become classical just because they stopped using data from one of the detectors.

Anwar and Ori also discussed how something could act quantum mechanically at a similar part of the second lab session. After seeing the two-detector measurement, Ori asked: ``If it's not behaving quantum mechanically, because the wave function hasn't been collapsed, it is able to get split up? Because it's a wave?'' Ori explained the role the experimenters took in this while summarizing the second experiment at the start of the third lab session. Although they accidentally switched which one was quantum and which one was classical, they said,
\begin{quote}
    The last [experiment] was totally mind blowing, and I thought about it all day. And basically, we saw that when we were measuring and observing the down-converted photons over there, they acted like classically over here. If we didn't, then they would act quantum mechanically.
\end{quote}
By choosing whether or not to account for observations of photons in all three detectors, the students affected whether the light was behaving quantum mechanically or classically.

The students discussed photons ``acting quantum mechanically'' even in contexts that did not take into account actions taken by the experimenters. When measuring the second-order correlation parameter while also looking at single-photon interference in the third experiment, Ori discussed how the light was ``acting quantum mechanically when it's acting like a wave.'' This came after the realization that the wave-like behavior of photons could be modeled by quantum mechanics. 

In most of these examples, the students saw a connection between the actions they as experimenters took and the observed results being classified as either quantum or classical. However, the language they sometimes used to discuss these ideas was language not typically used by experts.

\subsection{Discussion and instructional implications of student ideas about quantum versus classical}

The three themes presented in this section are quite different from each other, yet all represent various ways students discussed in their own words what it means to be quantum or classical. These themes encompass ontological, epistemological, and linguistic aspects, some of which may be tied together. The first theme, \emph{Waves and interference are classical}, relates to students' ontological understanding of waves and photons and the behavior each can exhibit. The theme \emph{Quantum mechanics is math} may be related to students' ontological classification of the theory itself, while also incorporating ideas of where this knowledge comes from. The last theme, \emph{Quantum can be turned off}, relates to the specific language the students used, but also connects to ontological reasoning about how different types of entities behave. 

Students already had some of these conceptions when arriving at the lab sessions, and others ideas formed or changed while the students worked with the experiments. Both groups of students initially talked about waves and interference as being classical, which may have come from prior physics or math courses or from the framing of the second experiment where students were introduced to the second-order correlation parameter. However, after seeing experimental evidence of single-photon interference, one of the groups used their knowledge of wave functions to realize that interference can also indicate that something is exhibiting quantum behavior. This opportunity to see experimental results that students had only previously learned about mathematically helped the students recognize that quantum mechanics is not just abstract math, it can also describe what happens in physical apparatus.

Student discussions about what was quantum versus classical occurred only in the second and third lab sessions. Although the first experiment involved the students measuring pairs of entangled photons, which are inherently quantum, we did not notice any such comments or conversations occurring at that time. This may be because a clear distinction between an object being quantum versus classical was first explicitly discussed in the second lab guide, where students were told there was an inequality that could be used to determine whether the light was best described by a classical or quantum model. 

It is therefore possible that the students' views reported here were influenced by the lab guides or the researcher acting as an instructor. Being prompted to take measurements within both the quantum and classical regimes of an inequality may have contributed to the students using language stating that quantum mechanics can be turned off or is a way that objects can behave. All of the student quotes related to this idea were discussed in reference to measurements of the second-order correlation parameter. None of the students brought up this idea when seeing other ways that an outcome is affected by measurements an experimenter could take, such as when performing the quantum eraser and thinking of which-path information. It is not clear if that is because students had already accepted that interference could be quantum, because the lab guide did not focus on a distinction between quantum or classical for that measurement, or for some other reason. Nonetheless, instructors should be aware of the language they use and the framing they provide, as their students may adopt the same phrasing.

These themes, and the way some student ideas surrounding them changed, demonstrate that experiments may provide an alternate and productive way for students to think about the distinction between quantum and classical models. The ideas that quantum mechanics is not purely math and that experimental choices affect measurement outcomes may be easier for students to internalize, and understand the nuances of, while working with physical experiments.
Quantum experiments may be especially helpful in allowing students to understand how quantum mechanics came about from successful predictions of experimental results.

\section{Conclusions} \label{sec:conclusions}

In this study, we performed three two-hour lab sessions with two pairs of students to understand their in-the-moment reasoning while making sense of experiments demonstrating particle-wave duality of photons. This allowed us to identify specific ideas the students came in with, and how some of those ideas changed while working through the sequence of experiments. 

We first identified several resources students activated while engaging with the idea that single photons can interfere with themselves. Both pairs of students activated various resources related to classical wave interference, one pair also activated several resources related to wave functions, and the other pair instead activated the resources \textit{Quantum outcomes are probabilistic} and \textit{Thinking in terms of information}. These resources and the ways students used them provide an example for instructors of the background knowledge students might need as they work with the single-photon experiments, as well as ways instructors may be able to help students build off of their already existing ideas.

We additionally investigated students' ideas about what is quantum or classical about these experiments and found both ontological and epistemological reasoning. The three primary ways this appeared in the data was when the students discussed that waves and interference are classical, that quantum mechanics is just math, and that quantum mechanics is something that can be turned on or off. Although these ideas appeared at various times throughout the second and third lab sessions, there was evidence that for some students these ideas transformed as they progressed through the experiments. Instructors should attend to the terminology they use to discuss the ideas in these experiments and also use the experiments as an opportunity to provide additional nuance to complex quantum concepts, such as the role experimenters play in obtaining possible outcomes.

We found that the complex topics exhibited in the single-photon experiments connected different aspects of students' understanding and allowed the students to productively engage with their own ideas in various ways. Other work has identified how some of the ``messiness'' in student reasoning surrounding quantum mechanics can be productive for students \cite{hoehn2018students}, and this is likely true in an experimental context as well. The quantum versus classical themes demonstrated how various types of student reasoning, as well as the language they used, all came into play while the students discussed conceptual topics. The students used some of their resources across multiple contexts, for example by using their knowledge of wave functions not only to understand their experimental results, but also to realize that interference is not just a classical concept, it can also be indicative of quantum behavior. Instructors could attend to the way different kinds of reasoning come together in quantum experiments and acknowledge the various productive ideas students bring to and generate while working with these experiments.

Another instructional implication related to the analyses of both research questions relates to the challenge of language in teaching students about, and assessing their comprehension of, quantum phenomena. Other studies have shown that the often ambiguous, context-dependent, and metaphorical language instructors use when teaching quantum mechanics can pose a difficulty for students' conceptual learning \cite{brookes2007using, marshman2015framework, bouchee2022towards} and that students often use the same language as their instructors \cite{serbin2021characterizations}. It is therefore unsurprising that ambiguous language also showed up in the student ideas we identified and may have contributed to differences between the resources activated by the two pairs of students. Luce even recognized the limitations of their own vocabulary in the post-interview by expressing that they wished they had acquired ``a more robust vocabulary'' to have better discussed and recorded their ideas during the lab sessions. Acknowledging this limitation may help both instructors and education researchers. Instructors could consider ways to help students connect their resources to more precise terminology and be aware that their own choice of words may be carried over to their students. Researchers may need to be careful in their assessment of student ideas, as it can be difficult to interpret student reasoning about nuanced topics when imprecise language is used.

This work highlights the importance of providing students opportunities to work with physical quantum experiments and the need for additional studies on the efficacy of this approach. Obtaining hands-on experience with quantum experiments is not only useful for students interested in entering the quantum workforce \cite{fox2020preparing, aiello2021achieving}, but it may also provide a different perspective for students learning about abstract or unintuitive quantum phenomena and how the mathematical theory can be motivated by experiments. However, there are many challenges to implementing a large-scale study of student conceptual learning gains from working with the single-photon experiments, including the limited number of students in upper-division lab courses and the difficulty of creating a validated assessment for experiments that are implemented in different ways and with different goals in each course. Nonetheless, there are many open questions to which the community should attend. Future work is needed to implement a large-scale study of conceptual learning with these experiments, to identify resources other students activate while working with these experiments and related ones (e.g., the Bell's inequality experiment), and to understand the role that lab partners or groups play as students reason through the seemingly strange experimental results that quantum mechanics predicts.

\begin{acknowledgements}
We would like to thank the students who participated in this study for dedicating their time and being enthusiastic participants. We also thank Mark Beck (via his published book), Kiko Galvez, David Hanneke, Amy Lytle, and Kevin Van De Bogart for sharing resources that helped with the development of our lab guides as well as Carlo Giacometti, Rachael Merritt, and Mike Bennett for providing feedback on the initial versions of the lab guides. Also thanks to the PER group at CU Boulder for useful conversations and feedback. This work is supported by NSF Grant PHY 2317149 and NSF QLCI Award OMA 2016244.
\end{acknowledgements}

\appendix

\section{Student conversations} \label{app:allConvos}

In this appendix, we present the longer conversations from the third lab sessions that show the detailed context in which the pairs of students activated the identified resources.

\subsection{Anwar and Ori 1} \label{app:AnwarOri1}

\linenumbers
\linenumbersep=4pt
\noindent \hangindent=.5cm \textbf{Anwar}: When there's more stuff hitting both A and B', there's less stuff hitting both A and B...

\noindent \hangindent=.5cm \textbf{Ori}: Yeah. I agree. Because when there's more going in here, there's less going in here. And vice versa. Just because the interference pattern.

\noindent \hangindent=.5cm \textbf{Anwar}: Because the interference pattern? 

\noindent \hangindent=.5cm \textbf{Ori}: Yeah. Because like, that's what we saw on the piece of paper.

\noindent \hangindent=.5cm \textbf{Anwar}: Yeah, that one of the beam paths would get brighter or less bright. That makes sense. 

\noindent \hangindent=.5cm \textbf{Ori}: Yeah.

\noindent \hangindent=.5cm \textbf{V.B.}: Is this what you were expecting to see?

\noindent \hangindent=.5cm \textbf{Anwar}: With a single photon, no. (\textit{Laughs.}) Right, this is what we saw visually with our eyes with the laser.

\noindent \hangindent=.5cm \textbf{Ori}: Yeah, oh I forgot that--- I was telling myself this was the laser so it would make sense. Shoot. 

\noindent \hangindent=.5cm \textbf{Anwar}: So, not really. No. (\textit{Laughs.}) It's surprising. 

\noindent \hangindent=.5cm \textbf{Ori}: It's--- Yeah, how does that even happen? Like, we were expecting that changing the mirror distance would do nothing. So why is it behaving classically?

\noindent \hangindent=.5cm \textbf{V.B.}: Is it necessarily behaving classically?

\noindent \hangindent=.5cm \textbf{Ori}: I guess not. I mean, we know that it's not.

\noindent \hangindent=.5cm \textbf{V.B.}: So what is necessary for something to interfere with it--- to interfere?

\noindent \hangindent=.5cm \textbf{Ori}: A wave? Well, yeah.

\noindent \hangindent=.5cm \textbf{Anwar}: Do you mean just like constructive or deconstructive effects? Waves stuff? 

\noindent \hangindent=.5cm \textbf{V.B.}: Yeah

\noindent \hangindent=.5cm \textbf{Anwar}: Two, two things.

\noindent \hangindent=.5cm \textbf{V.B.}: I guess. Yeah. So in this case you need two like wave-like things that somehow get put back into the same place.

\noindent \hangindent=.5cm \textbf{Anwar}: And are either going to constructively make the amplitude higher or deconstructively cancel each other out.

\noindent \hangindent=.5cm \textbf{V.B.}: So how--- maybe how could you then think about the photon at this first beam splitter differently if the way you were thinking about it is not leading to the results that you're seeing?

\noindent \hangindent=.5cm \textbf{Ori}: Maybe it's like changing--- is it changing the wave function? So it'll have different probabilities of going into each path.

\noindent \hangindent=.5cm \textbf{Anwar}: It's like the wave function of the photon. It's now like there's a probability of it going on each path. 

\noindent \hangindent=.5cm \textbf{Ori}: Yeah.

\noindent \hangindent=.5cm \textbf{Anwar}: And then by changing the mirror or the piezo---

\noindent \hangindent=.5cm \textbf{Ori}: I don't know how that would work. 

\noindent \hangindent=.5cm \textbf{Anwar}: ---we're changing how the wave function reflects.

\noindent \hangindent=.5cm \textbf{Ori}: I'm confused.

\noindent \hangindent=.5cm \textbf{V.B.}: So you can think of the piezo, the mirror it's still just changing the path length, but that's effectively like imparting a phase on any part of the wave function that is in this arm of the interferometer. 

\noindent \hangindent=.5cm \textbf{Anwar}: Okay. I'm okay with that.

\noindent \hangindent=.5cm \textbf{Ori}: Wait. That makes sense because like a wave function is time dependent. And so you're shifting---

\noindent \hangindent=.5cm \textbf{Anwar}: Well, I think it makes sense because the wave function is like spatial.

\noindent \hangindent=.5cm \textbf{Ori}: I mean, in either scenario. 

\noindent \hangindent=.5cm \textbf{Anwar}: Both. Okay yeah. Both. Both is good.

\noindent \hangindent=.5cm \textbf{Ori}: Huh. So when you--- That's crazy. Okay, so when you add a phase to the wave function, and it sort of interacts with each other in different ways. 

\noindent \hangindent=.5cm \textbf{Anwar}: Yeah, so the coh--- so there's--- so okay. Like, the interference we're seeing is just the deconstructive or constructive interference of the wave functions on each path. Which is like weird because it makes me want to think of the wave function as like a real---

\noindent \hangindent=.5cm \textbf{Ori}: Like an actual wave.

\noindent \hangindent=.5cm \textbf{Anwar}: ---thing. 

\noindent \hangindent=.5cm \textbf{Ori}: Yeah. Rather than, like, a set of determining probabilities.

\noindent \hangindent=.5cm \textbf{Anwar}: Yeah. But that makes sense because if the wave functions are like... a wave function kind of like probability distribution. If the wave functions are constructing with each other, then it's more likely that the thing will end up there. 

\noindent \hangindent=.5cm \textbf{Ori}: Yeah.

\noindent \hangindent=.5cm \textbf{Anwar}: So if we adjust the path lengths such that they're, the wave function constructs at a detector, it's more likely that the photon will be observed to be there. And if we have a bunch of them, then we're gonna see higher counts there. 

\noindent \hangindent=.5cm \textbf{Ori}: That makes sense. Yeah. 

\noindent \hangindent=.5cm \textbf{Anwar}: At a moment.

\noindent \hangindent=.5cm \textbf{Ori}: That's kind of crazy, though. 

\noindent \hangindent=.5cm \textbf{Anwar}: I liked it.

\noindent \hangindent=.5cm \textbf{Ori}: I like it. It would be cool to like do this side by side with like--- I, we totally did like math problems like this in our quantum class, where you like add a phase and see what happens.

\noindent ...

\noindent \hangindent=.5cm \textbf{Anwar}: Well, cool. All right. So it doesn't make immediate sense, but wave functions. Wild.

\nolinenumbers

\subsection{Anwar and Ori 2} \label{app:AnwarOri2}

\resetlinenumber
\linenumbers
\linenumbersep=4pt

\noindent \hangindent=.5cm \textbf{Ori}: No, that totally went up. For sure. 

\noindent \hangindent=.5cm \textbf{Anwar}: Okay.

\noindent \hangindent=.5cm \textbf{Ori}: And AB' went down a lot. Okay, because now we only have one beam. So you're dealing with a different wave function, instead of two. And they're not recombining. There's no phase.

\noindent \hangindent=.5cm \textbf{Anwar}: Hmm. Unblock this one, cause this, the effects are more dramatic. Okay. What's happening?

\noindent \hangindent=.5cm \textbf{Ori}: Okay, so we're looking at AB'. So it's the ones that were going there. I mean, you're removing the interference pattern. So---

\noindent \hangindent=.5cm \textbf{Anwar}: Yeah, we're removing the probability function on this side, right? So there is no, this probability distribution and, or this wave function and this wave function interfering anymore.

\noindent...

\noindent \hangindent=.5cm \textbf{Anwar}: I still don't understand why it's going that way.

\noindent \hangindent=.5cm \textbf{Ori}: Well, because it's returning to the original wave function. 

\noindent \hangindent=.5cm \textbf{Anwar}: What do you mean?

\noindent \hangindent=.5cm \textbf{Ori}: Like, instead of having to deal with the destructive interference coming from the phase-changed wave function. Now, it's just one pure thing. And so it's like more favored probability-wise towards AB. Like, the only reason this was really low is because of the interference. And that's kind of what we're proving.

\noindent \hangindent=.5cm \textbf{Anwar}: The only reason this was low because of the interference between these two beam paths? 

\noindent \hangindent=.5cm \textbf{Ori}: Yeah.

\noindent \hangindent=.5cm \textbf{Anwar}: This one and this one? I didn't understand that.

\noindent \hangindent=.5cm \textbf{Ori}: Yeah, like this is at a minimum of that, like, sinusoid right now. 

\noindent \hangindent=.5cm \textbf{Anwar}: Mmhmm.

\noindent \hangindent=.5cm \textbf{Ori}: And because it's at a m--- it's at a minimum because of the interference that's happening between the two wave functions. And when you remove one of the paths, you don't have that interference anymore. And so it just sort of returns to like, what it would be if it was just one wave function.

\noindent \hangindent=.5cm \textbf{Anwar}: Oh, I see. 

\noindent \hangindent=.5cm \textbf{Ori}: And it happens that that increases the number of counts.

\noindent \hangindent=.5cm \textbf{Anwar}: Oh, I see. Okay, so I think what--- is this what you're saying?  Like I'm just going to try to rephrase it--- 

\noindent \hangindent=.5cm \textbf{Ori}: Yeah.

\noindent \hangindent=.5cm \textbf{Anwar}: --- so it makes sense to me. Like. The probability, or the wave functions, right now, when there's two paths, they are like constructively, well, they're deconstructing each other right now, right? 

\noindent \hangindent=.5cm \textbf{Ori}: Yeah.

\noindent \hangindent=.5cm \textbf{Anwar}: That's why there is a small probability of there being a, we're looking at AB'? 

\noindent \hangindent=.5cm \textbf{Ori}: AB.

\noindent \hangindent=.5cm \textbf{Anwar}: AB. There's a small probability of there being a photon on detector B, right? 

\noindent \hangindent=.5cm \textbf{Ori}: Yeah.

\noindent \hangindent=.5cm \textbf{Anwar}: Because the probabilities from this path, or the wave function to this path and this path are deconstructing---

\noindent \hangindent=.5cm \textbf{Ori}: Yeah.

\noindent \hangindent=.5cm \textbf{Anwar}: ---the probabilities are going to be low. And because the probability of a photon being here is low, the coincidence with A is also being very low. 

\noindent \hangindent=.5cm \textbf{Ori}: Yeah.

\noindent \hangindent=.5cm \textbf{Anwar}: But then when we get rid of one, it's not deconstructing anymore. 

\noindent \hangindent=.5cm \textbf{Ori}: Right. 

\noindent \hangindent=.5cm \textbf{Anwar}: So there's the higher probability that it's going to be in B because there's a higher probability that it's going to be in B.

\noindent \hangindent=.5cm \textbf{Ori}: AB.

\noindent \hangindent=.5cm \textbf{Anwar}: There's a higher probability there's a coincidence between A and B.

\noindent \hangindent=.5cm \textbf{Ori}: That's how I'm interpreting it.

\noindent \hangindent=.5cm \textbf{Anwar}: I like that. I like that. That took me a while to get there, I don't know why, but that makes sense. 

\noindent \hangindent=.5cm \textbf{Ori}: No, that's okay. It's weird.

\nolinenumbers

\subsection{Anwar and Ori 3} \label{app:AnwarOri3}

\resetlinenumber
\linenumbers
\linenumbersep=4pt

\noindent \hangindent=.5cm \textbf{Ori}: And so we're seeing that it's really close to zero, which means that it's acting quantum mechanically. Oh. So we're saying it's acting quantum mechanically when it's acting like a wave because it's a wave function. Like, there's interference. So we're saying---

\noindent \hangindent=.5cm \textbf{Anwar}: So this $g^{(2)}$ measurement says that it is either at one place or the other, right? Because there's very low coherence.

\noindent \hangindent=.5cm \textbf{Ori}: Yeah, there's very few coincidence counts in ABB', like in all three of them at the same time.

\noindent \hangindent=.5cm \textbf{Anwar}: Yeah. So that sounds like it's either at one place or the other place. 

\noindent \hangindent=.5cm \textbf{Ori}: Right.

\noindent \hangindent=.5cm \textbf{Anwar}: But what we just talked about with the interference pattern is that it's best understood as a wave function.

\noindent \hangindent=.5cm \textbf{Ori}: Which makes sense?

\noindent \hangindent=.5cm \textbf{Anwar}: Yeah. Because the probability of it being at one place, or B, meant that it was very low probability of being at B', right?

\noindent \hangindent=.5cm \textbf{Ori}: Yeah.  Like the wave functions is acting like a wave, but it's really just determining the probability of the particle going into one or the other.

\noindent \hangindent=.5cm \textbf{Anwar}: Right, so I think it's okay that it looks like it's at one place or the other.

\noindent \hangindent=.5cm \textbf{Ori}: Yeah. Well, cause it is.

\noindent \hangindent=.5cm ...

\noindent \hangindent=.5cm \textbf{V.B.}: Does it all kind of make sense the way--- what is happening to the photon at this beam splitter and then at that beam splitter? 

\noindent \hangindent=.5cm \textbf{Ori}: Yeah. I feel pretty comfortable with this.

\noindent \hangindent=.5cm \textbf{Anwar}: I do too. I like it... It's weird to think of things as like a sp--- I'm thinking of it spatially at least, as like a spatial wave function across the breadboard. In fact, it makes me want to like measure it, like here, here, here, here, here, here, here and then, like, measure the wave function.

\nolinenumbers

\subsection{Kiran and Luce 1} \label{app:KiranLuce1}

\resetlinenumber
\linenumbers
\linenumbersep=4pt

\noindent \hangindent=.5cm \textbf{Kiran}: And the $g^{(2)}$ is where it was before more or less... 0.09 which is our much-less-than-one quantum-y case.

\noindent \hangindent=.5cm \textbf{V.B.}: So what is that telling you?

\noindent \hangindent=.5cm \textbf{Luce}: That we are getting single photons.

\noindent \hangindent=.5cm \textbf{V.B.}: What's happening to the single photons at the beam splitter?

\noindent \hangindent=.5cm \textbf{Luce}: They are remaining single photons. Right? [Kiran]?

\noindent \hangindent=.5cm \textbf{Kiran}: The coincidences are not happening at the same time. Which is telling us the single photon effect, right?

\noindent \hangindent=.5cm \textbf{Luce}: Yeah.

\noindent \hangindent=.5cm \textbf{V.B.}: And what's happening at this beam splitter? 

\noindent \hangindent=.5cm \textbf{Luce}: Huh. It's a wave. It's a laser. Laser (\textit{points}). Laser. Photon.

\noindent \hangindent=.5cm \textbf{V.B.}: Is the laser going on that path (\emph{points to entire interferometer}) at all?

\noindent \hangindent=.5cm \textbf{Luce}: On this path?

\noindent \hangindent=.5cm \textbf{V.B.}: Uhhuh.

\noindent \hangindent=.5cm \textbf{Luce}: No. Right, [Kiran]?

\noindent \hangindent=.5cm \textbf{Kiran}: Which path are you---

\noindent \hangindent=.5cm \textbf{Luce}: No laser. Maybe laser? No laser.

\noindent \hangindent=.5cm \textbf{Kiran}: I mean, it's coming off of that and then bouncing in the detectors.

\noindent \hangindent=.5cm \textbf{Luce}:  Yeah. No laser. But as single photons! We're detecting single photons---

\noindent \hangindent=.5cm \textbf{Kiran}: Yeah. 

\noindent \hangindent=.5cm \textbf{Luce}: ---in this path. And could be lasers over there.

\noindent \hangindent=.5cm \textbf{V.B.}: But yeah, what does that mean that you're seeing interference also?

\noindent \hangindent=.5cm \textbf{Luce}: If we're seeing interference, it means there has to be waves in here. But we're measuring single photons here. So somehow the waves are going from waves to single photons here.

\noindent \hangindent=.5cm \textbf{Kiran}:  Or could it be that the single photons were waves all along?

\noindent \hangindent=.5cm \textbf{Luce}: Oh my god. 
Dun dun dun.

\noindent \hangindent=.5cm \textbf{V.B.}: So what is a way to think about like if you do have a single photon that's coming here. What like, how can you talk about what happens to it right here?

\noindent \hangindent=.5cm \textbf{Luce}: You can't talk about what happens in the beam splitter to a single photon. Because then you'd have half a photon and that's not a thing.

\noindent \hangindent=.5cm \textbf{Kiran}: But you can talk about what might happen.

\noindent \hangindent=.5cm \textbf{Luce}: It could go one way or the other, if it was a single photon.

\noindent \hangindent=.5cm \textbf{Kiran}: Two-dimensional Hilbert space.

\noindent \hangindent=.5cm \textbf{Luce}: But if it was a wave, it could go both ways. Right?

\noindent \hangindent=.5cm \textbf{Kiran}: Such are the properties of---

\noindent \hangindent=.5cm \textbf{Luce}: Which is why we're seeing interference because it's going both ways. 

\noindent \hangindent=.5cm \textbf{Kiran}: ---waves.

\noindent \hangindent=.5cm \textbf{Luce}: If it was a single photon, it would basically be like you're blocking one of the arms. So it would just like not go in one of the arms.

\nolinenumbers

\subsection{Kiran and Luce 2} \label{app:KiranLuce2}

\resetlinenumber
\linenumbers
\linenumbersep=4pt

\noindent \hangindent=.5cm \textbf{Luce}: The polarizer is realigning it so that it can interfere again. And so we get interference again. But it's after the interferometer.

\noindent \hangindent=.5cm \textbf{Kiran}: Well, the polarizer is destroying the information of how the---

\noindent \hangindent=.5cm \textbf{Luce}: The non-interfered. 

\noindent \hangindent=.5cm \textbf{Kiran}: ---surviving photons are polarized.

\noindent \hangindent=.5cm \textbf{Luce}: It's destroying the non-interfered photons. So we're only getting the interfered photons. Right? We're erasing the non-interfered photons.

\noindent \hangindent=.5cm \textbf{V.B.}: When you say interfered photons or non-interfered photons, what do you mean by that?

\noindent \hangindent=.5cm \textbf{Luce}: The photons that were--- I don't know. I don't know what I'm saying. So photons come in here. They get shifted, shifted, so they don't interfere and they don't interfere. Here, they're shifted. And then we only get the ones that get shifted back.

\noindent \hangindent=.5cm \textbf{V.B.}: I mean, it's--- like the polarizer's still gonna be a probabilistic process, right? 

\noindent \hangindent=.5cm \textbf{Luce}: Yeah. 

\noindent \hangindent=.5cm \textbf{V.B.}: For every photon here, it kind of projects it back onto the state and makes it through. 

\noindent \hangindent=.5cm \textbf{Luce}: Yeah. 

\noindent \hangindent=.5cm \textbf{V.B.}: Like it could project on this state or it could project onto the orthogonal state that just gets absorbed instead. 

\noindent \hangindent=.5cm \textbf{Luce}: Yeah. 

\noindent \hangindent=.5cm \textbf{V.B.}: So all the photons then have a 50\% chance of going through. 

\noindent \hangindent=.5cm \textbf{Luce}: Yes. 

\noindent \hangindent=.5cm \textbf{V.B.}: And at that point, could you tell which way, which arm of the interferometer it had gone through?

\noindent \hangindent=.5cm \textbf{Luce}: Yes.

\noindent \hangindent=.5cm \textbf{V.B.}: How?

\noindent \hangindent=.5cm \textbf{Luce}: That's a good fucking question. (\emph{Everyone laughs.}) [Kiran], do you have an answer to that question? You haven't talked that much...

\noindent \hangindent=.5cm \textbf{Kiran}:  I think that you can't tell. But I also can't super justify that. Except to say that if you were able to tell, if you were able to say it went one way, you would know, if you receive it and you see its polarization, and you know how it started, you know which waveplate rotated it. But if there's the polarizer in the way, just information-wise, you don't--- everything you receive is the same polarization and it could have come from one or the other, or classically the wave. Well, classically it's gone forever. But. You know, to translate the information argument into physics argument is not simple.

\noindent \hangindent=.5cm \textbf{Luce}: The polarizer's erasing information. So then we don't know which one it came from?

\noindent \hangindent=.5cm \textbf{Kiran}: And therefore, since it could have come from either, it can do the interference.

\noindent \hangindent=.5cm \textbf{Luce}: Yes, because it's back in a superposition. So these ones (\emph{points to non-interfering counts on computer screen}) are being more particle-y. And these ones are being more wave-y?

\nolinenumbers

\bibliography{SinglePhotonClinicalStudy}

\clearpage


\section*{Supplementary material}

# Student reasoning about quantum mechanics while working with physical experiments: Supplemental material

In this supplement, we provide additional information about the lab guides used by the students during the three lab sessions in this study.

As discussed briefly in Sec. III A of the main paper, these materials were pieced together from and inspired by many sources, including work from Mark Beck, Kiko Galvez, David Hanneke, Amy Lytle, and Kevin Van De Bogart. Because some of their materials are already published, we are not able to provide the complete lab guides here. However, we summarize the background information provided in these guides (Sec. I) and also include the complete procedure sections (Secs. II-IV). Please contact the authors if you are interested in the complete lab guides.

## I. SUMMARY OF LAB GUIDES

The students were provided five different documents during the three lab sessions. At the start of the first lab session, the students received a document about personal and lab safety that focused on actions the students must follow to ensure proper laser safety. It also explained the procedures necessary to prevent the equipment, in particular the single-photon detectors and optical components, from being damaged.

After a discussion about lab safety, the students were provided a document with background information relevant for all three experiments. This contained a brief summary of the sequence of experiments; an overview of the experimental apparatus (with a diagram similar to, but more detailed than, those of Fig. 1 in the main paper); and descriptions of spontaneous parametric down-conversion, the single-photon detectors, coincidence counting, and the specific LabVIEW program the students would be using. This document also included some references students could look up if they wanted more detailed explanations of the theoretical backing of the experiments. Because the background information relevant for the first experiment was relevant for the later two as well, the procedure for the first experiment (see Sec.II) was provided to the students separately.

At the start of the second lab session, the students were provided a single lab guide for the day including background information specific to that experiment and the procedure. The background section of this lab guide provided a theoretical description of photon correlations, including a description of the second-order correlation parameter $g^{(2)}(0)$ for both two- and three-detector measurements. It also had a section about the new additions to the apparatus, including a short description of beam splitters and a diagram of the layout of the experiment with the new additions of a beam splitter, a third detector, and the alignment laser. The procedure section from the lab guide for the second experiment is shown in Sec. III.

For the third lab session, the students were again provided a single lab guide containing both the new background material and the procedure. The background section discussed interference, both with classical waves and single photons, and included a diagram of a Mach-Zehnder interferometer, as well as a photo of the apparatus with the different parts of the interferometer and the piezoelectric actuator (piezo) marked. This was followed by information about piezos, including a very brief description of how piezos function in general, as well as specific information about how the students could control the voltages sent to the piezo in this specific experiment. The procedure section from the lab guide for the third experiment is shown in Sec. IV.

## II.  PROCEDURE FOR LAB 1 – SPONTANEOUS PARAMETRIC DOWN-CONVERSION

In the first lab of this sequence, you will demonstrate that you can generate pairs of photons and distinguish between these and background fluctuations. You will begin by looking at the different parts of the apparatus and making sure you understand each part. You will then optimize and align the fiber couplers that send the pairs of photons from the optical table to the detectors to help you understand the different experimental knobs that are available. Lastly, you will record numbers of counts of photons and use these to convince yourself that this experimental setup will allow you to perform experiments with single photon states.

The purpose of this entire sequence of experiments is to see phenomena and understand qualitatively what is happening. Although you will be asked to record numbers of counts of photons or perform small calculations with them, there is no single right or wrong answer. Unless any question in this set of lab manuals specifically asks you for a numerical value with its error, a by-eye average will be sufficient. If, at a later point, you realize you need to be more precise with a measurement, you can always go back and spend more time taking data when it is needed.

The procedure is below. Don't hesitate to ask if you have any questions.

1. **Preparing the detectors to record photon counts**

    (a) Before beginning the experiment, point out each of the following to your lab partner: 405 nm laser, half-waveplate, BBO crystal, detectors, and fiber couplers labeled A and B'. Show each other the path of the down-converted photons. If you have any questions about the experimental set-up, please ask the researcher before proceeding.

    (b) Turn on the two laser warning signs (labeled switches next to room light switches by door).

    (c) Turn off the room lights (and make sure the window is covered). You may want to use a headlamp or phone light to walk around the room while the lights are off as long as you are careful not to shine it near the detectors.

    (d) Open `Coincidences.vi` in LabVIEW 2014. Run the program by clicking on the white arrow just below the `Edit` menu. You should see counts of 0 for everything since the detectors have not yet been turned on.

    (e) While looking at the counts in LabVIEW (to ensure they never go above 1 million counts/second), turn on the power supply for detectors A and B' by flipping first the big switch (that will light up red/orange when it's on) and then the smaller switches for each individual detectors (green LEDs will be illuminated for each detector when on).

    (f) Once detectors A and B' are on, you should be measuring only dark counts since the laser has not yet been turned on. Estimate the number of dark counts for detectors A and B' and the AB' coincidence counts, and record those estimates in your lab notebook. What is the time window over which these counts are measured? Do the dark counts change as you swivel the computer screen? Does anything else affect the dark counts?

2. **Optimizing alignment of fiber couplers to down-converted photons**

    (a) Make sure everyone in the room is wearing the correct laser safety goggles and the laser warning signs outside the lab have been turned on.

    (b) Turn on the 405 nm laser and ensure you can see it passing through the crystal.

    (c) Detector A should already be approximately optimized, but you will finalize this optimization, first by adjusting the vertical tilt of the crystal. Look at the counts on detector A while turning the top knob (that controls the vertical tilt) on the mount of the crystal by a small fraction of a turn. Be careful to keep your hand out of the path of the laser as you do this. (Note that the crystal is most sensitive to vertical tilts because only vertically polarized light will be down-converted.) After slightly rotating the knob in both directions until you see a decrease in counts on detector A, position the vertical knob such that it maximizes the counts.

    (d) Continue optimizing the number of counts on detector A by adjusting one at a time the horizontal and vertical knobs on the mirror sending light into the fiber coupler for detector A followed by the horizontal and vertical tilts of the fiber coupler for detector A. Rotate each knob gradually, leaving it at a position that maximizes the counts. Record the number of counts on detector A in your lab notebook.

    (e) At this point, you will hopefully see a large number of counts on detector A, but you need to make sure these counts are actually down-conversion and not just background. Rotate the half waveplate just after the 405 nm laser to ensure that the counts depend on the polarization of the laser. Leave the half waveplate at an angle that maximizes the number of counts on detector A.

(f) The next step is to place the fiber coupler for detector B' such that it maximizes the coincidence counts AB'. Since detector A is 32 inches away from the crystal, place detector B' so it is also 32 inches away from the crystal and roughly lined up with the path of the other down-converted photons. (Hint: the holes on the optical table are each 1 inch apart.) Make slight adjustments to the position and angle to see if you can see any coincidence counts. Once you've found the position that approximately maximizes the coincidence counts AB', screw the fiber coupler down to the table.

(g) Optimize the coincidence counts AB' by adjusting one-by-one first the two knobs on the mirror immediately before the fiber coupler for detector B' and then the two knobs on the fiber coupler itself. Record the number of counts on A, B' and AB' in your notebook. What fraction of the total counts are coincidence counts?

3. **Understanding the down-converted photons**

   (a) One way to test whether the coincidence counts you are detecting are coincidences created by down-conversion or whether they are accidental coincidences (counts that just happen to be detected within the same time window) is to add an intentional delay between the two signals. Turn off the detectors and the FPGA (the red button is the on/off switch for the FPGA), and then replace the short BNC cable between detector B' and the FPGA with a ten-meter-long BNC cable. Turn the FPGA and detectors back on (while looking at the counts in LabVIEW to ensure the detectors are never receiving too much light). What happens to the coincidence counts AB'? Is this what you would expect? Discuss with your lab partner.

   (b) Compare the number of coincidence counts you were measuring in step 2g to the number of dark counts measured in step 1f and accidental coincidence counts measured in step 3a. Can you argue that the noise is a small fraction of the signal?

   (c) Down-converted photons are produced at random times, but suppose that on average they are evenly separated in time. Using the measured coincidence counts (taking into account dark counts and accidental coincidences), how far apart in space on average are the down-converted photons? Is it reasonable to say that there's only one photon interacting with the experimental setup at a time?

   (d) Once you have taken all the measurements you want to take, turn off the 405 nm laser and the detectors. You can then take off the laser safety goggles and turn on the room lights. The researcher will ensure everything else is turned off.

# III. PROCEDURE FOR LAB 2 – EXISTENCE OF A PHOTON

In Lab 1, you examined the behavior of the SPDC source and learned how to maximize the coincidence count rate between detectors A and B'. In this lab, you will start from the end of Lab 1, where the alignment of detectors A and B' has already been completed. Your task in this lab will be to insert a beamsplitter in the path of the photons traveling towards B', and align a third detector B such that you can measure $g^{(2)}(0)$. You can then compare a two-detector measurement of $g^{(2)}(0)$ with a three-detector measurement to understand the purpose of using coincidence measurements.

1. **Alignment of beamsplitter and detector B**

    (a) Turn on the laser warning signs (or ensure they are already on).

    (b) The only new optics that have been added since your last lab are ones for the alignment laser, marked with red stickers. This laser has been aligned to go along the same path as one of the down-converted photons to help you know where to place the beamsplitter and third detector. Point out the path the alignment laser takes to your labmate. After that, find the beamsplitter and fiber coupler for detector B that you will be adding to the setup, so you know where they are once the lights are out.

    (c) Make sure the fiber coupler to detector B' is fully blocked with a sheet of paper or card. (Be careful of the optical fibers nearby when moving around parts near the fiber couplers, since the fibers can be relatively easily damaged.) Put down the flip mirror that sends the photon towards B' after the crystal. Turn on the alignment laser.

    (d) Insert a beamsplitter around 3 inches away from the B' fiber coupler. Make sure the alignment laser passes approximately through the center of it and that the reflected beam is centered on the line of holes on the breadboard around 3 inches away (since that is where the fiber coupler will go). Once the beam splitter is in place, screw it to the table.

    (e) Turn off the alignment laser, block it with a card, and flip up the flip mirror after the crystal. Once that is done, go ahead and unblock fiber coupler B'.

    (f) Place the mount with the fiber coupler for detector B on the reflected side of the beamsplitter around 3 inches away from the beamsplitter, centered on the line of holes where you saw the alignment laser reflected by the beamsplitter. Put the screws in, but don't tighten them yet.

    (g) Turn off the room lights. Open and run the LabVIEW file `Coincidences.vi`. While looking at the counts in LabVIEW (to ensure they don't go above 1 million counts/second), turn on detectors A, B, and B'.

    (h) Make sure everyone is wearing laser safety goggles. Then turn on the 405 nm laser.

    (i) Adjust the position of fiber coupler B by moving it slightly side to side and rotating it slightly until you have roughly maximized the AB coincidence counts by hand. Be careful to keep your hand out of the path of the 405 nm laser. Don't worry if the counts are not as high as you'd expect; you will be able to increase these counts more in the next step. Now tighten the screws so the fiber coupler is firmly mounted on the table.

    (j) One-at-a-time, adjust two of the knobs on the beamsplitter mount (the top one and the one on the bottom closest to it) followed by the vertical and horizontal tilt of the B fiber coupler to optimize AB coincidence counts. (Hint: if you block detector B' while doing this, the graph in LabVIEW will scale with AB coincidence counts making it easier to see small changes.)

    (k) Record in your lab notebook the AB and AB' coincidence counts as well as the A counts. Are the coincidence counts what you'd expect based on the AB' coincidence counts you were getting in Lab 1? If not, discuss with your labmate what could be causing this discrepancy.

2. **Measurement of $g^{(2)}(0)$ with a single photon state**

    (a) Close the current LabVIEW file and open the LabVIEW file `g2measurements.vi`. What new features are in this program? Discuss them with your labmate.

    (b) Rotate the knob to `g(2) 3-det`. What physically does the number of ABB' three-fold coincidence counts represent? Discuss with your labmate and record the number of ABB' counts in your lab notebook.

    (c) In the pane labeled `g(2) Measurements`, click `Clear Buffer` and watch the results that come in afterward to get an idea of the average value of $g^{(2)}(0)$. You can see the value of $g^{(2)}(0)$ from each run plotted on the `g(2)(0)` tab of the pane with the plots.

(d) Take data (around 10 points in order to ensure reasonable statistics) with an update period of anywhere between 0.1 and 10 seconds. To do this, click the `Take Data` button in the `Data Taking Parameters` box after setting the update period and number of points for the data run. A new window will pop up as it's taking the data. The data file will be saved to `C:\Lab\Data\2023\[today's_date]`. Note that the main LabVIEW file stops after saving data so you'll have to run it again with the white arrow below the `Edit` menu to have it continue to record counts. Record the average value of $g^{(2)}(0)$ along with the standard deviation of the measurements in your lab notebook.

(e) Take another set of data with a different value of the update period (still between 0.1 and 10 seconds). Record the average value of $g^{(2)}(0)$ along with the standard deviation in your lab notebook and compare it with the value from your previous measurement. Do they agree?

(f) Do your measured values of $g^{(2)}(0)$ agree with the expected value (to within the error of the measurement)? Discuss with your labmate why you shouldn't measure a value of exactly 0 for $g^{(2)}(0)$.

3. **Measurement of two-detector $g^{(2)}(0)$**

Now that you've hopefully convinced yourself that single photons act like particles at beamsplitters, let's see what happens when we remove the conditioning of the measurement on the counts at detector A. You will now show that a single beam of your down-conversion source behaves classically when you ignore the other beam.

(a) Running the same LabVIEW file (`g2measurements.vi`), turn the knob to `g(2) 2-det`. You should see that the triple coincidence counts ABB' are now replaced by two detector coincidence counts BB'. Write down the number of BB' coincidence counts in your notebook.

(b) Click `Clear buffer` and look at the data collected afterward to get an estimate for the two detector measurement of $g^{(2)}(0)$.

(c) Take one set of data (similar to the three detector measurement). Record the average value of $g^{(2)}(0)$ with its standard deviation in your lab notebook.

(d) Discuss with your labmate why your values for $g^{(2)}(0)$ differ when performing the two-detector and three-detector measurements. What does this say about what is needed to experimentally ensure you are working with a single photon state?

(e) Once you have taken all the measurements you want to take, turn off the 405 nm laser and the detectors. You can then take off the laser safety goggles and turn on the room lights. The researcher will ensure everything else is turned off.

## IV. PROCEDURE FOR LAB 3 – SINGLE PHOTON INTERFERENCE

1. **Interference with laser light**

   (a) Look at the apparatus and identify each new piece that has been added since your last lab. The blue stickers are placed on optical elements that are part of the Mach-Zehnder interferometer (and as before, the red stickers are placed on optical elements used for the alignment laser). With your lab partner, indicate the paths the photons will take through the interferometer.

   (b) With your lab partner, discuss what you would expect to see at both outputs of the second beamsplitter in the interferometer when you send 1) a laser and 2) single photons into the interferometer. How will you know if you are seeing interference?

   (c) Block (e.g., with a sheet of paper) the fiber couplers going to detectors B and B' and ensure the detectors are still powered off. Then, turn on the HeNe alignment laser (make sure the laser warning signs are on first!) and flip down the flip mirror so the alignment laser can reach the interferometer. What do you see at the outputs of the interferometer? Can you see changes in intensity in the light by tapping the table or blocking one arm of the interferometer? Discuss with your lab partner what is happening and if it matches what you'd expect.

   (d) Open the piezo controller GUI Kinesis on the computer. Without going out of the range of the piezo $(0-100$ V$)$, vary the voltage while looking at the output of the second beamsplitter of the interferometer. What do you see happening? Discuss with your lab partner.

   (e) Turn off the HeNe alignment laser, place the Thorlabs card immediately after the first mirror in its path to block the alignment laser, flip up the flip mirror that sends the down-converted photons into the interferometer, and remove the paper blocking detectors B and B'.

2. **Interference with single photons**
   Now that you have seen a laser interfere with itself, let's see what happens when you send single photons (the states you prepared and measured in the last two labs) into the interferometer.

   (a) Open and run the LabVIEW file `g2measurements.vi` (with the knob turned to `Coincidence`).

   (b) Double check the alignment laser has been fully blocked (see step 1e). Turn off the room lights. Turn on detectors A, B, and B' while watching the LabVIEW file to ensure the counts never go above 1 million counts/second.

   (c) Make sure everyone puts on laser safety goggles. Turn on the 405 nm laser.

   (d) Record the number of counts on detectors B and B' as well as the coincidence counts AB and AB' in your lab notebook.

   (e) Block one arm of the interferometer (a good way to do this is by holding a business card immediately after one output of the first beamsplitter) and again record the counts B, B', AB, and AB'. Then record these counts while blocking the other arm of the interferometer. Do these numbers match what you'd expect? Discuss with your lab partner.

   (f) Using Kinesis, vary the voltage sent to the piezo (keeping it within the range of $0-100$ V), while looking at the counts AB and AB'. What happens? Draw a sketch of the coincidence counts vs. voltage in your notebook.

   (g) Find a piezo setting such that you are at a minimum for the coincidence counts AB. Block one of the arms of the interferometer. What do you observe happening to the coincidence counts AB? Does it make sense? Now go to a maximum in the coincidence counts AB (and therefore a minimum in the counts AB'). Again, what do you observe? Discuss these observations with your partner and write down a brief explanation in your lab notebook.

   (h) Turn the knob in the LabVIEW program to `g(2) 3-det`. Measure $g^{(2)}(0)$ (either by cleaning the buffer and looking in the main window or using the data taking function) and record that value along with its standard deviation in your lab notebook. Can you reconcile your interference measurement with the measurement of $g^{(2)}(0)$? Discuss with your lab partner.

   You have now demonstrated both particle-like and wave-like behaviors of light in the same experiment!

3. **Quantum eraser**
   To finish off this sequence, you will perform a short follow-up where you will look at the effect of "marking"

the photons to indicate which arm of the interferometer they go through. To do this, you will use the half-waveplates and polarizer (marked with yellow stickers). Before beginning this section, discuss with your lab partner what happens to linearly polarized light when it passes through a half-waveplate and when it passes through a polarizer.

(a) Turn off the 405 nm laser and the detectors.

(b) Physically block (e.g., with paper) the fiber couplers for B and B' to protect them from the next step.

(c) Flip down the flip mirror, unblock the alignment laser, and turn on the alignment laser once you have double checked that the detectors are both off and blocked.

(d) While being very careful not to bump the interferometer, insert the two half-waveplates, putting one in each arm of the interferometer (the researcher can show you where they fit). You will want to align them so that the alignment laser passes through approximately the center and the small amount of light reflected from each half-waveplate goes back on the incoming beam. An easy way to check this is to look at the unused port of the first beamslitter in the interferometer.

(e) Make sure that both half-waveplates are rotated to the number labeled on the top as being the axis (one should be at 16 degrees and the other at 52 degrees). When they are aligned such that the light is parallel to one of the axes of the waveplate, the waveplates will not rotate the polarization of the incoming light.

(f) Turn off and block the alignment laser and then flip up the flip mirror.

(g) Once you have double checked the alignment laser is off and blocked, unblock the detectors and turn them back on.

(h) Make sure everyone has on laser safety goggles and turn back on the 405 nm laser.

(i) Ramp the piezo over the same range you were looking at before. Do you still see interference?

(j) Rotate one of the half-waveplates by 45 degrees (this corresponds to a rotation in polarization of 90 degrees, therefore changing the light into an orthogonal polarization). Ramp the piezo again. What happened to the interference? Discuss what you are seeing with your lab partner.

(k) Now place the polarizer (that is set at 45 degrees to that of the photons in the interferometer) after the interferometer but before the fiber coupler to detector B'. Again ramp the piezo. What happens to the AB' coincidence counts? Why might this be called a quantum eraser? What is the difference between the AB coincidence counts and the AB' coincidence counts? Discuss with your lab partner.

(l) If there's anything else you want to try out or play with involving the set-up, discuss it with the researcher. Once you are done, turn off the 405 nm laser and the detectors. You can then take off the laser safety goggles and turn on the room lights. Return the piezo to 0 V.



\end{document}